\documentclass[twocolumn,times]{aastex62}

\usepackage[version=4]{mhchem}
\setcitestyle{number}
\usepackage{comment}

\newcommand{\change}[1]{{#1}}

\newcommand{\ee}[1]{\ensuremath{\times 10^{#1}}}

\submitjournal{ApJ}

\begin{document}
    
    \title{Ab initio Study of Ground-State CS Photodissociation Via Highly Excited Electronic States} 
    
    \correspondingauthor{Kyle N. Crabtree}
    \email{kncrabtree@ucdavis.edu}

    \author{Zhongxing Xu}
    \affiliation{Department of Chemistry, University of California – Davis, One Shields Avenue, Davis, CA 95616, USA}

    \author{Nan Luo}
    \affiliation{Department of Chemical Engineering, University of California – Davis, One Shields Avenue, Davis, CA 95616, USA}

    \author{S. R. Federman}
    \affiliation{Department of Physics and Astronomy, University of Toledo, Toledo, OH 43606, USA}

    \author{William M. Jackson}
    \affiliation{Department of Chemistry, University of California – Davis, One Shields Avenue, Davis, CA 95616, USA}

    \author{Cheuk-Yiu Ng}
    \affiliation{Department of Chemistry, University of California – Davis, One Shields Avenue, Davis, CA 95616, USA}

    \author{Lee-Ping Wang}
    \affiliation{Department of Chemistry, University of California – Davis, One Shields Avenue, Davis, CA 95616, USA}

    \author{Kyle N. Crabtree}
    \affiliation{Department of Chemistry, University of California – Davis, One Shields Avenue, Davis, CA 95616, USA}

    \begin{abstract}
    Photodissociation by ultraviolet radiation is the key destruction pathway for CS in photon-dominated regions, such as diffuse clouds.
    However, the large uncertainties of photodissociation cross sections and rates of CS, resulting from a lack of both laboratory experiments and theoretical calculations, limit the accuracy of calculated abundances of S-bearing molecules by modern astrochemical models.   
    Here we show a detailed \textit{ab initio} study of CS photodissociation.
    Accurate potential energy curves of CS electronic states were obtained by choosing an active space CAS(8,10) in MRCI+Q/aug-cc-pV(5+d)Z calculation with additional diffuse functions, with a focus on the \(B\) and \(C\,^1\Sigma^+\) states.
    Cross sections for both direct photodissociation and predissociation  from the vibronic ground state were calculated by applying the coupled-channel method. 
    We found that the \(C-X\) \((0-0)\) transition has extremely strong absorption due to a large transition dipole moment in the Franck-Condon region and the upper state is resonant with several triplet states via spin-orbit couplings, resulting in predissociation to the main atomic products C \((^3P)\) and S \((^1D)\).
    Our new calculations show the photodissociation rate under the standard interstellar radiation field is \(2.9\ee{-9}\)\,s\(^{-1}\), with a 57\% contribution from \(C-X\) \((0-0)\) transition.
    This value is larger than that adopted by the Leiden photodissociation and photoionization database by a factor of 3.0.
    Our accurate \textit{ab initio} calculations will allow more secure determination of S-bearing molecules in astrochemical models. 

    \end{abstract}

    \keywords{molecular data --- molecular processes --- interstellar chemistry}
    
    \section{introduction}\label{sec:intro}
    Sulfur is an abundant element in space, e.g., the relative abundance of S to H is 1.3\(\times\)10\(^{-5}\) in the solar system \citep{Asplund2009}, and the abundances of S-bearing molecules are sensitive to the physical conditions of their environments.
    In the interstellar medium (ISM), S-bearing molecules are commonly detected and used as tracers of physical properties \citep{Semenov2018}.
    In star forming regions, it has been suggested that abundances of \ce{H2S}, SO and \ce{SO2} could act as a chemical clock on the time scale of \(10^4\) years due to both thermal heating and shock interactions \citep{Tak2003,Wakelam2011}.
    In protoplanetary disks, the abundances of S-bearing species may correlate with the C/O ratio, surface diffusivity, turbulent mixing, X-ray luminosity, ultraviolet (UV) intensity and grain growth \citep{Semenov2018}.

    However, the abundances of S-bearing species are poorly reproduced by modern astrochemical models \citep{Lucas2002}, possibly because of the large uncertainties in kinetic data, missing reaction pathways, and unaccounted reservoirs of sulfur \citep{Druard2012,Loison2012,Vidal2017}. 
    CS was the first sulfur-bearing molecule observed in interstellar space, initially detected by its 3--2 rotational emission line at 146.969\,GHz in several dense sources \citep{Penzias1971}. 
    It has since been found in a wide set of diffuse and dense interstellar clouds \citep{Zuckerman1972,Drdla1989,Heithausen1998,McQuinn2002,Scappini2007}, as well as comets \citep{Jackson1982, Canaves2007}.
    Additionally, CS is the key species in the sulfur chemistry of protoplanetary disks.
    Observations of the CS column density are used to determine upper limits for other S-bearing molecules since CS is the only detected sulfur species in many disks, such as DM Tau \citep{Semenov2018}. 

    In photon-dominated or photodissociation regions (PDRs), UV photons play a critical role in the gas phase chemistry and act as the most important source of energy. 
    In a general sense, PDRs include peripheries of molecular clouds, diffuse clouds, translucent clouds, the surfaces of protoplanetary disks, and cometary and exoplanetary atmospheres. 
    For small molecules like CS, photodissociation is the key destruction pathway in those environments. 
    Accurate chemical modeling requires the wavelength-dependent photoabsorption/photodissociation cross sections at energies above the dissociation limit.
    
    While the ground \(X\,^1\Sigma^+\) and several low-lying electronic states (\(a\,^3\Pi\), \(a^\prime\,^3\Sigma^+\), \(d\,^3\Delta\), \(e\,^3\Sigma^-\), \(A\,^1\Pi\) and \(A^\prime\,^1\Sigma^+\)) of CS have been extensively studied by both experiments and \textit{ab initio} calculations \citep{Shi2013}, very few studies have been done on highly excited states in vacuum UV (VUV) region where CS may undergo photodissociation.
    The pioneering study on highly excited states of CS was by  \citet{Crawford1934}, who assigned a strong band system around 251\,nm to CS in the emission spectrum of a low-pressure discharge of \ce{CS2}.
    Later \citet{Donovan1970} recorded the first VUV spectrum of CS via time-resolved flash photolysis of \ce{CS2} coupled with a high-resolution spectrograph.
    A strong band observed at 154.1\,nm was assigned as \(B\,^1\Sigma^+-X\,^1\Sigma^+\) by analogy with the valence isoelectronic species CO, which suggests the \(B\) state of CS has a Rydberg nature like the corresponding state of CO.
    Two more strong bands at 140.2 and 139.9\,nm were assigned as the \(C\,^1\Sigma^+-X\,^1\Sigma^+\) \((0-0)\) and \((1-1)\) transitions, also by analogy with CO.
    
    A subsequent high-resolution VUV absorption study of CS by \citet{Stark1987} confirmed the \(C-X\) band assignment and also found additional vibrational components of the \(B-X\) transition.
    Their rotational contour analysis of the \(B-X\) \((1-0)\) band found that the spectroscopic constants of \(B\) state are close to those of the \ce{CS+} ground state, strongly supporting the proposed Rydberg nature of the \(B\) state.
    A rough measurement showed that the linewidth of the \((1-0)\) band is on the order of 1\,cm\(^{-1}\), which is clearly broadened by predissociation. 
    All other bands were too diffuse to show rotational structures.
    Both the \(C-X\) \((0-0)\) and \((1-1)\) bands were diffuse and intense, indicating the Franck-Condon factors of this transition must notably favor the \((0-0)\) transition.
    The experimental assignments were supported by an early SCF-CI calculation \citep{Bruna1975}, which found that the \(B\) and \(C\) Rydberg states agreed with experimental energies within 0.1\,eV.
    The spectroscopic evidence suggests that the \(B-X\) and \(C-X\) bands should play important roles in CS photodissociation in space owing to their strong intensities and their broadening by predissociation.
    However, at present the best estimates of the CS photodissociation cross sections in the Leiden database \citep{Heays2017} were made by combining the measured \(B-X\) transition wavelength and vertical excitation energies of higher valence and Rydberg states, and are estimated to be uncertain to a factor of 10.

    To improve the accuracy of photodissociation data for astronomical models, further experiments and high-level quantum chemical calculations are needed.
    Most recently, \citet{Pattillo2018} performed the first high-level \textit{ab initio} calculations targeting states involved in CS photodissociation.
    They concluded that the dominant contribution to CS photodissociation from the ground electronic state comes from direct excitation of several dissociative states, including \(A'\,^1\Sigma^+\) and several \(^1\Pi\) states, while predissociation via the \(B\) state is unimportant.
    However, their results show significant discrepancies with the experimental VUV spectroscopy of the \(^1\Sigma^+\) states: specifically, the energy of the \(B\) state is about 7000\,cm\(^{-1}\) higher than the experimental value and the shape of its potential energy curve indicates a much lower vibrational constant compared with experiments, and the \(C\) state is missing entirely.
    Thus, the conclusion that predissociation in highly excited states is unimportant should be re-examined more carefully.

    Here, we present a high-level \textit{ab initio} study of CS photodissociation, including for the first time a detailed investigation of its predissociation via the \(B\,^1\Sigma^+\) and \(C\,^1\Sigma^+\) states. We found that under the Draine radiation field \citep{Draine1978}, inclusion of the \(C-X\) and \(B-X\) transitions increases the CS photodissociation rate by nearly an order of magnitude compared with the results of \citet{Pattillo2018}, and yields an overall rate that is higher by a factor of 3 compared with the Leiden database~\citep{Heays2017}.
    The details of our theoretical methods are introduced in Section~\ref{sec:theory}. 
    The computed potential energy curves, transition dipole moments, photodissociation cross sections, and photodissociation rates are discussed in Section~\ref{sec:dissc}, as well as the comparison between our calculations and experiments. 
    Finally, a summary of the work and its future directions are given in Section~\ref{sec:conclu}.
    
    \section{theory and calculations}\label{sec:theory}
    
    \subsection{Ab initio calculation}\label{subsec:theory_abinitio}

    Our calculations use the state-averaged complete active space self-consistent field (SA-CASSCF) approach \citep{Werner1985,Knowles1985}, followed by internally contracted multireference configuration interaction with single and double excitations and the Davidson correction (MRCI+Q) \citep{Werner1988,Knowles1988,Knowles1992}, a widely used method for calculating excited electronic states, especially for diatomic molecules. 
    We used the the quantum chemical package MOLPRO 2015.1 \citep{Werner2012,Werner2015} to calculate the adiabatic potential energy curves (PECs) and transition dipole moments (TDMs) of CS.
    
    To determine the PECs accurately, up to a total of 105 single point calculations with internuclear separation between 0.78 to 7.93\,\AA\ were carried out, with step sizes ranging from 0.0026 to 0.26\,\AA. 
    The smaller step sizes were used near the equilibrium geometry of the ground state and in the vicinity of several important avoided crossings between states with the same symmetry to ensure good accuracy of calculated properties. 
    We used Dunning's augmented correlation consistent polarized valence quintuple-zeta Gaussian basis set with tight \(d\) orbitals for sulfur [aug-cc-pV(5+d)Z or AV(5+d)z]\citep{Kendall1992,Dunning2001}. 
    The tight \(d\) orbitals have been shown to be essential for calculating accurate properties of S-bearing species \citep{Trabelsi2018}. 
    Several additional diffuse Gaussian functions corresponding to Rydberg atomic orbitals (AOs) of C and S were added to the basis sets to more accurately represent the Rydberg character of the \(B\) and \(C\) states. 
    Their exponents, derived from \citet{Schaefer1977}, are given in Table~\ref{tab:basis_set}. 

    \begin{deluxetable*}{CCCCCCCCC}[hbtp]
        \tablewidth{\linewidth}
        \tablecaption{Exponents of diffuse Gaussian functions added to the aug-cc-pV(5+d)Z basis set\label{tab:basis_set}}
        \tablehead{
              \multicolumn5c{C} & &\multicolumn3c{S} \\ \cline{1-5} \cline{7-9}
              \colhead{\(3s\)} & \colhead{\(3p\)} &\colhead{\(3d\)} &\colhead{\(4s\)} &\colhead{\(4p\)} & &\colhead{\(3d\)} &\colhead{\(4s\)} &\colhead{\(4p\)}
        } 
        \startdata
        0.01725 & 0.01575 & 0.02850 & 0.01045 & 0.00931 & & 0.02850 & \change{0.01725} & 0.02949  \\
                &         & 0.01125 & 0.00413 & 0.00368 & & 0.01125 &         & 0.01500 
        \enddata
    \end{deluxetable*}

    MOLPRO is unable to take advantage of the full symmetry of non-Abelian groups (in this case, C\(_{\infty v}\)), so the calculation is performed in the largest Abelian subgroup (C\(_{2v}\)).
    The reducing map of irreducible representations from C\(_{\infty v}\) to C\(_{2v}\) is \(\Sigma^+\rightarrow A_1\), \(\Sigma^-\rightarrow A_2\) \(\Pi\rightarrow (B_1,B_2)\), and \(\Delta\rightarrow (A_1,A_2)\). 
    We adopt MOLPRO's order of irreducible representations for C\(_{2v}\) to indicate the number of molecular orbitals (MOs) of each symmetry in the following discussion, \((a_1, b_1, b_2, a_2)\).

\begin{figure}[hbtp!]
    \centering 
    \includegraphics[width=0.42\columnwidth]{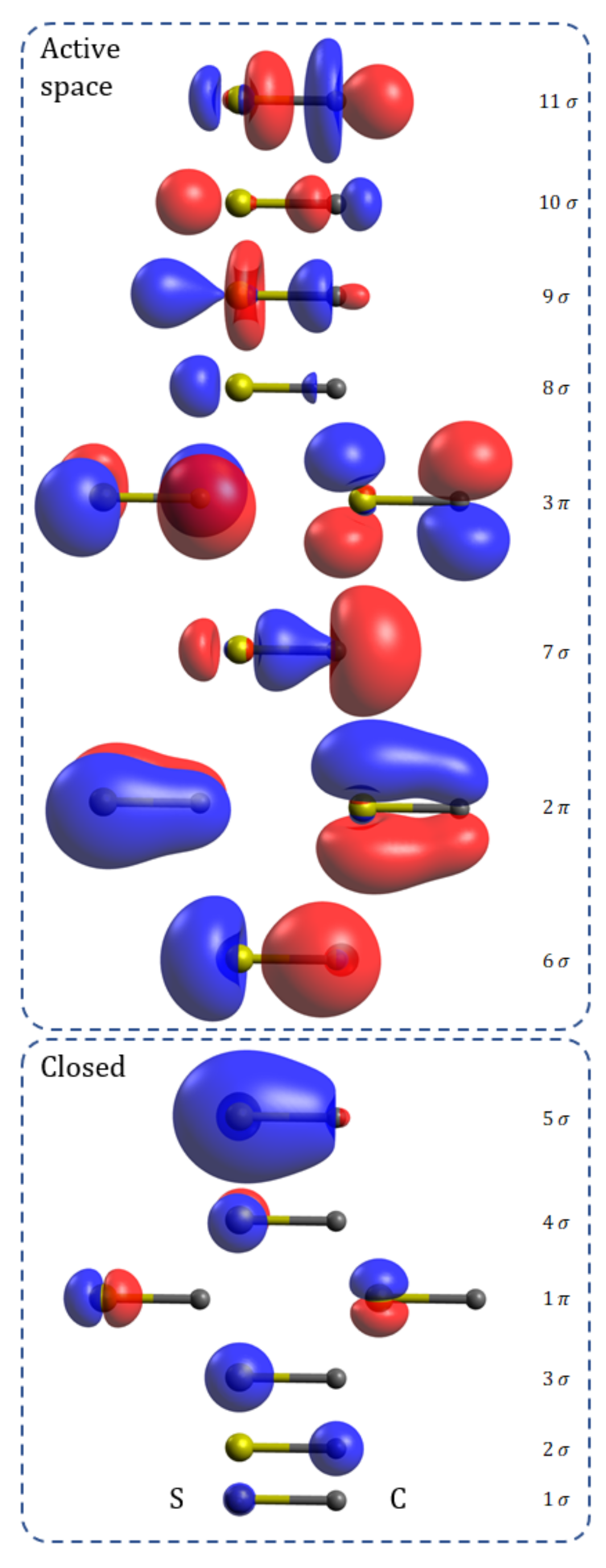}
    \caption{Molecular orbitals (MOs) of CS calculated by SA-CASSCF at 1.54\,\AA, plotted with isovalue 0.08, except 0.02 is used for the \(1\,\sigma\) MO. Contributions from \(h\) type orbitals are excluded from this image owing to limitations of the visualization software. In $C_{2v}$, \(\sigma\) orbitals correspond to \(a_1\) and \(\pi\) orbitals to the pair \((b_1,b_2)\). The MOs are listed according to increasing energy, though not to scale. The orbital shapes depend strongly on internuclear distance.}\label{fig:MOs}
\end{figure}

    The dominant electron configuration of CS in its ground (\(X\,^1\Sigma^+\)) state at its equilibrium geometry is \(1\sigma^2\)\(2\sigma^2\)\(3\sigma^2\)\(4\sigma^2\) \(5\sigma^2\)\(6\sigma^2\)\(7\sigma^2\)\(1\pi^4\)\(2\pi^4\).
    To construct the active space for our SA-CASSCF/MRCI+Q calculation, 17 MOs (11,3,3,0) were involved in total.
    The 7 MOs (5,1,1,0) with lowest energies are kept closed (doubly-occupied) in the reference space, while the remaining 8 electrons are distributed in the other 10 MOs, forming an active space CAS(8,10) (6,2,2,0).
    A more detailed discussion on our choice of active space is given in Section~\ref{subsubsec:dissc_abinitio_activespace}.
    The MOs included in our calculation described above are shown near the equilibrium geometry of the ground state in Figure~\ref{fig:MOs}.

    The ground states of atomic carbon and sulfur are both \(^3P\), followed by an excited state \(^1D\). 
    The lowest four dissociation limits of CS therefore correlate to C (\(^3P_{0,1,2}\)) + S (\(^3P_{0,1,2}\)), C (\(^3P_{0,1,2}\)) + S (\(^1D\)), C (\(^1D\)) + S (\(^3P_{0,1,2}\)), and C (\(^1D\)) + S (\(^1D\)). 
    These combinations give rise to 82 electronic states in C\(_{2v}\) symmetry, which are 10 \(^1A_1\), 8 \(^1B_1\), 8 \(^1B_2\), 8 \(^1A_2\), 9 \(^3A_1\), 10 \(^3B_1\), 10 \(^3B_2\), 10 \(^3A_2\), 3 \(^5A_1\), 2 \(^5B_1\), 2 \(^5B_2\), and 2 \(^5A_2\) states.
    We carried out the averaging process among the 82 lowest-energy states of these C\(_{2v}\) symmetries in the SA-CASSCF calculations regardless of which symmetries they correspond to in C\(_{\infty v}\); this changes with internuclear distance.

    The orbitals optimized by SA-CASSCF are used in the MRCI+Q calculations. 
    The CI treatment was carried out by employing a reference space of 2053 (\(^1A_1\)), 1843 (\(^3A_1\)) 1672 (\(^1B_1\) and \(^3B_1\)), 1368 (\(^1A_2\) and \(^3A_2\)), 891 (\(^5A_1\)), 1000 (\(^5B_1\)), and 1144 (\(^5A_2\)) configurations, from which all single and double excitations were generated.
    As a demonstration of the calculation size, the total number of uncontracted configurations was 502\,853\,808 while the total number of contracted configurations was 17\,569\,850 in \(^1A_1\) symmetry for the MRCI+Q calculation at 1.54\,\AA. 
    The exact number of configurations varies with internuclear distance, spatial symmetry, and spin multiplicity.
    The Davidson correction with relaxed references are added to the MRCI energies.
    TDMs are obtained at the level of MRCI.

    For several bound states with obvious potential wells, we calculated the spectroscopic constants from the PECs for comparison with previous calculations and experimental data. 
    First, we obtained the rovibrational energy levels by solving the one-dimensional Schr\"{o}dinger equation numerically using the DUO program \citep{Yurchenko2016}. 
    Then the spectroscopic constants, including \(T_e\), \(\omega_e\), \(\omega_ex_e\), \(B_e\), and \(\alpha_e\) were determined by fitting the energy levels of the first ten vibrational states.

    \subsection{Photodissociation cross sections}\label{subsec:theory_pd_cross_section}
    Photodissociation may occur through one of two main pathways.
    Absorption into an unbound excited electronic state results in direct dissociation, and is characterized by a broad, weak cross section.
    Indirect photodissociation on the other hand begins by absorption into a bound excited state, followed by predissociation: non-radiative coupling into a nearby unbound state.
    Cross sections for indirect photodissociation show resolved or partially-resolved rovibrational transitions associated with the upper electronic state that are lifetime broadened.
    When the predissociation timescale is fast compared with other relaxation mechanisms (e.g., spontaneous emission), nearly every absorption event leads to dissociation.

    The coupled-channel Schr\"{o}dinger equation (CSE) technique is employed here to study the predissociation mechanisms of CS \(^1\Sigma^+\) states. 
    When solving the Schr\"{o}dinger equation, there are two ways to describe the coupled system of nuclei and electrons.
    The electronic states calculated by the \textit{ab initio} methods above are in the adiabatic representation, where the electronic Hamiltonian is diagonalized precisely and the couplings between states arise from a nuclear kinetic energy operator.
    An alternative method uses the diabatic representation, where the nuclear kinetic energy coupling terms are minimized while introducing new couplings that are treated as interactions between different electronic states.
    The diabatic states approximately follow the same electronic character as a function of internuclear distance, while the adiabatic states' electronic character varies.
    In principle, these two representations are equivalent after introducing the appropriate coupling terms. 
    For convenience, the diabatic representation is used in this study because the couplings in adiabatic states vary strongly with internuclear distance, which creates difficulties in modeling the predissociation process.

    In the CSE method, the complete coupled-wavefunction \(\psi_i(\textbf{r},R)\) is expressed as a linear combination of a set of \(N_T\) diabatic (or adiabatic) electronic rotational states \(\phi_j(\textbf{r};R)\), which are also called coupled channels
    \begin{equation}\label{equ:coupled_states}
    \psi_i(\textbf{r},R) = \sum_{j=1}^{N_T}\frac{1}{R}\chi_{ij}(R)\phi_j(\textbf{r};R)
    \end{equation} 
    where \(R\) is the internuclear distance, and \(\chi_{ij}(R)\) represent \(R\)-dependent expansion coefficients between states \(\phi_j(\textbf{r};R)\).	
    For a given energy \(E\), the Schr\"{o}dinger equation for the radial wavefunctions can be written as 
    \begin{equation}\label{equ:CSE_equation}
        \frac{\partial^2}{\partial R^2}\chi(R) = -\frac{2\mu}{\hbar^2}\chi(R)[E\textbf{I}-\textbf{V}(R)]
    \end{equation}
    where \(\mu\) is the reduced mass of the molecule, \(\textbf{I}\) is the identity matrix, and \(\textbf{V}(R)\) is the interaction matrix, which is composed of potential energy curves as diagonal elements and coupling terms (such as non-adiabatic coupling and spin-orbit coupling) as off-diagonal elements.

    The spin-orbit couplings and non-adiabatic couplings are calculated by MOLPRO.
    The spin-orbit couplings for MRCI wavefunctions are calculated by using the full Breit-Pauli operator between internal configurations while contributions of external configurations are calculated by a mean-field one-electron Fock operator.
    For adiabatic states, the non-adiabatic coupling matrix elements (NACMEs) are computed by finite differences of the MRCI wavefunctions.  
    Details about building the interaction matrix, including obtaining diabatic representations, will be discussed further in Section~\ref{subsec:dissc_CSE}.
    Equation (\ref{equ:CSE_equation}) is solved numerically to give the coupled wavefunctions for mixed upper states. 
    
    Assuming alternate decay pathways such as spontaneous emission or collisional relaxation are slow, the total photodissociation cross section from an initial state with \(J^{\prime\prime}\) is obtained by summing over all open channels \(\gamma\) and all allowed \(J^{\prime}\) \citep{Heays2010}
    \begin{equation}\label{equ:cross_section}
        \sigma_{g}(\tilde{\nu}) = \sum_{J^{\prime}} \sum_{\gamma} \Big[\frac{\pi\tilde{\nu}}{3\hbar\varepsilon_0} \frac{g}{2J^{\prime\prime}+1} \sum_k( |\langle \chi_{\gamma k} | M | \chi_g \rangle|^2 S_{J^{\prime\prime}}^{\Delta J}) \Big]
    \end{equation}
    where \(\tilde{\nu}\) is the photon energy in wavenumbers, \(M\) is the R-dependent electric-dipole transition moment between the unmixed lower (ground) state with radial wavefunction \(\chi_g\) and each upper state \(k\) with mixed wavefunction \(\chi_{\gamma k}\) coupled to open channel \(\gamma\) . The H\"{o}nl-London factors \(S_{J^{\prime\prime}}^{\Delta J}\) \citep{Hansson2005,Watson2008} which indicate the relationship between the total intensity of a vibronic band and the rotational quantum numbers can be expressed for these types of transitions as:
    \begin{align}
    ^1\Sigma^+-^1\Sigma^+: \quad                    & S_{J^{\prime\prime}}^{P}            = J^{\prime\prime} \nonumber        \\
                         & S_{J^{\prime\prime}}^{R} = J^{\prime\prime}+1                  \\
    ^1\Pi-^1\Sigma^+     : \quad                    & S_{J^{\prime\prime}}^{P}            = (J^{\prime\prime}-1)/2 \nonumber  \\
                         & S_{J^{\prime\prime}}^{Q} = (2J^{\prime\prime}+1)/2  \nonumber  \\
                         & S_{J^{\prime\prime}}^{R} = (J^{\prime\prime}+2)/2 
    \end{align}
    for \(P (\Delta J = -1)\), \(Q (\Delta J = 0)\), and \(R (\Delta J = +1)\) branches. In our case, the degeneracy factor \(g\) is 1 for a \(^1\Sigma^+-^1\Sigma^+\) transition and 2 for a \(^1\Pi-^1\Sigma^+\) transition.

    For a particular transition, the linewidth can be used to estimate the predissociation timescale \(\tau_{pd}\) and compared with the spontaneous emission and collision timescales (\(\tau_{se}\) and \(\tau_{coll}\)).
    If \(\tau_{pd} << \tau_{se}\) and \(\tau_{coll}\), then the calculated cross sections are good estimates of the photodissociation cross section.
    Otherwise, a time-dependent method should be applied or a tunneling probability \(\eta\) should be included for correction. As shown below, in the case of CS, the predissociation efficiency is essentially 1.

    Direct photodissociation is simply a special case of the CSE model in which only one unmixed upper state can be excited from the ground state.
    Because the upper state is unbound and certain to dissociate, the calculated result is an exact photodissociation cross section.
    Thus, the CSE approach simultaneously calculates the direct photodissociation cross sections in addition to those that proceed via predissociation. 
    In this study, photodissociation cross sections are calculated with PyDiatomic \citep{Gibson2016}, which solves the time-independent coupled-channel Schr\"{o}dinger equation using the Johnson renormalized Numerov method \citep{Johnson1978}. 
    
    Using the CSE method, a rotationless (\(J^\prime-J^{\prime\prime}=0-0\)) transition is calculated for the ground \(X\) state with \(v^{\prime\prime}=0\).
    We also calculated the photodissociation cross sections for transitions from the ground state with \(v^{\prime\prime}=0,1,2\) and different \(J^{\prime\prime}\).
    Assuming local thermodynamic equilibrium (LTE), the total photodissociation cross sections at given temperature \(T\) are calculated by
    \begin{equation}\label{equ:lte_cs}
    \sigma(\lambda,T) = \frac{1}{Q}\sum_{i} \sigma_{i}(\lambda) g_i e^{{-{E_i}/{k_bT}}}
    \end{equation}
    where \(Q\) is the partition function, \(E_i\) is the energy of all achievable ground rovibrational states with rotational degeneracy \(g_i=2J^{\prime\prime}+1\), and \(k_B\) is Boltzmann's constant.
    
    \subsection{Photodissociation rates in astrophysical environments}\label{subsec:theory_pd_rate}
    The photodissociation rate of a molecule in an UV radiation field is
    \begin{equation}
    k = \int \sigma(\lambda) I(\lambda) d\lambda
    \end{equation}
    where \(\sigma(\lambda)\) is the photodissociation cross section and \(I(\lambda)\) is the radiation intensity.
    We compute the photodissociation rate of CS from its ground (\(X\)) state with \((v^{\prime\prime}, J^{\prime\prime}) =(0,0)\) in the standard interstellar radiation field (ISRF) given by \citep{Draine1978}.
    The LTE photodissociation rates for different temperatures are also calculated. 
    
    \section{Results and discussion}\label{sec:dissc}
    The layout of this section is as follows. The PECs and TDMs obtained from \textit{ab initio} calculations are shown in Section~\ref{subsec:dissc_abinitio}, including a highlight on the main feature of our calculations. Then, the details about building the coupled-channel model is discussed in Section~\ref{subsec:dissc_CSE}. Finally, the dissociation cross sections and rates are presented in Section~\ref{subsec:dissc_cross_sections}.

    \subsection{Ab initio calculation}\label{subsec:dissc_abinitio}

    \subsubsection{Optimization of MRCI calculation}\label{subsubsec:dissc_abinitio_activespace}
    
    The accuracy of the calculated photodissociation cross sections relies on the PECs and TDMs obtained from the SA-CASSCF/MRCI+Q calculation.
    The quality of an MRCI+Q calculation is sensitive to the choice of active space and basis set, both of which require careful consideration.
    Previous theoretical studies of CS excited states \citep{Shi2013,Pattillo2018} used the aug-cc-pV6Z (AV6Z) basis set with the active space CAS(10,8) where the number of active orbitals for each irreducible representation is given as (4,2,2,0).
    The fact that the properties of the \(B\) state calculated by \citet{Pattillo2018} disagree with experiments \citep{Donovan1970,Stark1987} suggests this active space is not suitable for accurately calculating highly excited states.
    One reasonable explanation for the discrepancy is that some dominant configurations of the \(B\) state are not included in the reference space because some significantly occupied MOs in those configurations are outside of the active space.
    
    Although there is no golden rule to determine the ideal active space, including more virtual orbitals is generally necessary to improve the quality of the calculation, especially for Rydberg states.
    Both previous spectroscopic experiments and comparison between CO and CS indicate the \(B\) and \(C\) states have Rydberg nature, involving high-energy \(\sigma\) type orbitals.
    Motivated by these experimental observations, we systematically included more \(a_1\) (i.e., \(\sigma)\) virtual orbitals into the active space, and found that at CAS(10,11) (7,2,2,0) the SA-CASSCF/MRCI+Q calculation was stable over the whole internuclear distance range.
    Smaller active spaces resulted in a stability problem around 2.0\,\AA.

    As the internuclear distance increases, the dominant electron configuration changes in the adiabatic representation.
    For the ground \(X\,^1\Sigma^+\) state, this occurs twice, at 2.1 and 2.8\,\AA, which can roughly be interpreted as the points at which the \ce{C=S} double bond breaks stepwise.
    While the change in configuration itself is straightforward to treat, the changes in the shapes of the MOs themselves causes significant stability problems when the active space is too small. 
    With our active space, we were able to achieve continuous and smooth PECs up to at least the \(C\,^1\Sigma^+\) state.
    Addition of one more \(\sigma\) orbital resulted in a dramatic increase in the single-point calculation time, rendering it impractical for the complete study.
    
    Calculations with smaller basis sets showed that the 5\,\(a_1\) MO is doubly occupied in the most important configurations for all states we are able to calculate.
    Therefore to save calculation time, we put the 5\,\(a_1\) MO into the closed-shell space, resulting in our final active space of CAS(8,10) (6,2,2,0).

    Because of our large active space, we could not use the aug-cc-pV6Z basis set as in previous studies.
    Instead, we used the aug-cc-pV(5+d)Z basis set supplemented with additional diffuse orbitals located on both carbon and sulfur atoms.
    The total number of AOs in our basis amounts to 299 (112,72,72,43).
    Keeping the 6 lowest MOs (4,1,1,0) as core MOs, in the MRCI+Q calculation for \(^1\Sigma^+\) states at 1.54\,\AA, from the reference space consisting of 2053 configurations, 1.76\,\(\times 10^{7}\) contracted and 5.03\,\(\times 10^{8}\) uncontracted configurations are generated.
    In comparison, in the aug-cc-pV6Z basis set there are 382 (134,93,93,62) AOs.
    To compute the same number of states using the active space CAS(10,8) (4,2,2,0) and the aug-cc-pV6Z basis set, only a total of 1.11\,\(\times 10^{7}\) contracted and 7.00\,\(\times 10^{7}\) uncontracted configurations are produced from the reference space with 240 configurations.
    Thus, our large reference space is appropriate for calculating both valence and Rydberg states of CS, and justifies using a slightly smaller, tailored basis set.

    As a final point, our choice of active space was focused primarily on accurate calculations of \(^1\Sigma^+\) states.
    It is possible that including more \(\pi\) MOs into the active space, such as using CAS(8,12) (6,3,3,0), would improve the quality of calculation especially for high-lying \(\Pi\) states.
    However, the large number of configurations we included in the MRCI+Q calculation still promises good accuracy even for non \(^1\Sigma^+\) states.
    Moreover, the spectroscopic constants calculated for low-lying excited states from our PECs match well with experiments where data are available, which enhances our confidence.
    
    \subsubsection{PECs and TDMs}\label{subsubsec:dissc_abinitio_pecs}
\begin{figure*}[hbtp]
        \includegraphics[width=\textwidth]{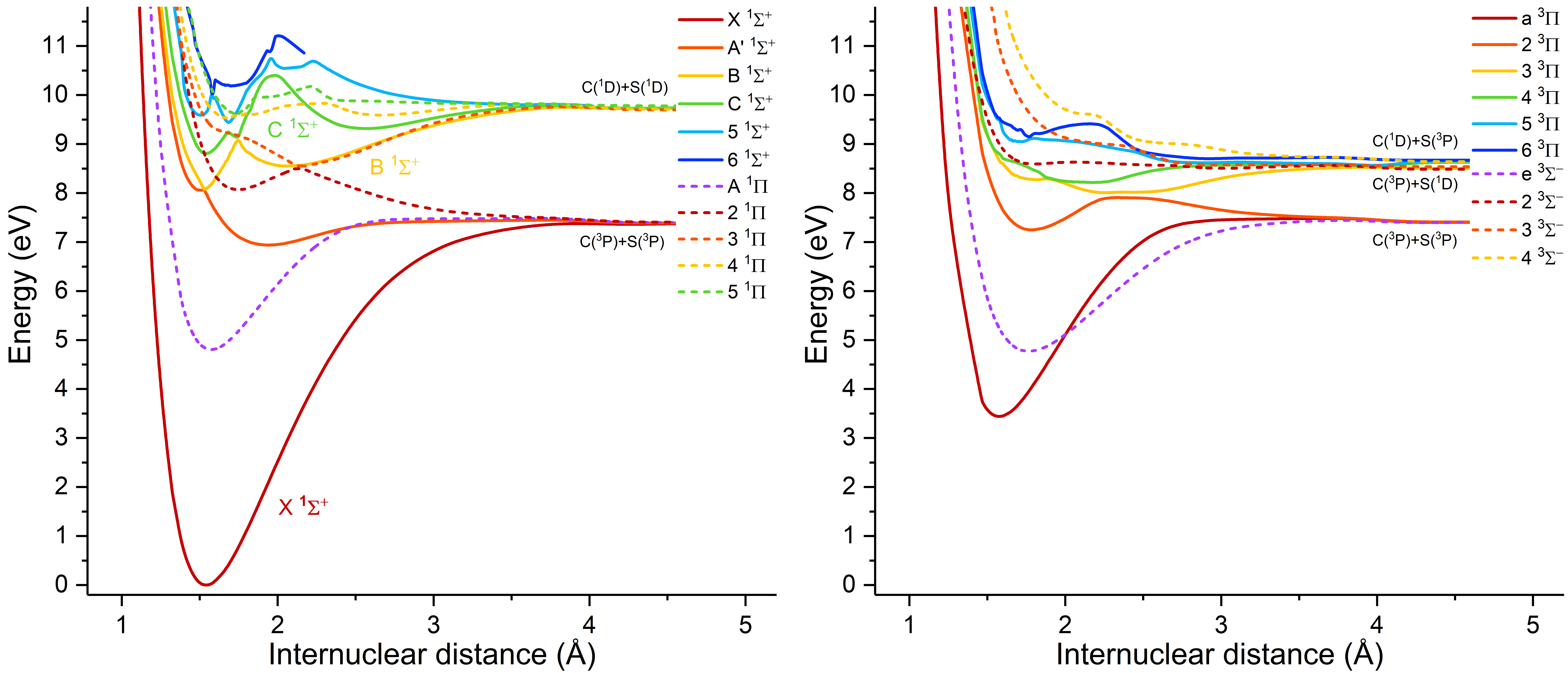}
        \caption{Adiabatic potential energy curves for CS \(^1\Sigma^+\), \(^1\Pi\) (left), \(^3\Pi\), and \(^3\Sigma^-\) (right) electronic states}\label{fig:PECs}
    \end{figure*}

    Employing the approaches described in Sec~\ref{subsec:theory_abinitio}, we have calculated the PECs of 49 states in total, including 7 \(^1\Sigma^+\), 3 \(^1\Sigma^-\), 7 \(^1\Pi\), 4 \(^1\Delta\), 4 \(^3\Sigma^+\), 5 \(^3\Sigma^-\), 8 \(^3\Pi\), 5 \(^3\Delta\), 2 \(^5\Sigma^+\), 1 \(^5\Sigma^-\), 2 \(^5\Pi\), and 1 \(^5\Delta\).
    Among all those states, the adiabatic PECs of several \(^1\Sigma^+\), \(^1\Pi\), \(^3\Pi\), and \(^3\Sigma^-\) states are shown in Figure~\ref{fig:PECs} because they are directly related to the following dissociation study, while all data are available in a machine-readable format in the Appendix with PECs of other states.

    The potential energy scale used here is referenced to a zero defined by the potential minimum of the ground state \(X\,^1\Sigma^+\). 
    State names are kept consistent for states already tabulated in the NIST database \citep{Huber1979}. 
    For the ground state and several low-lying excited states, calculated spectroscopic constants are listed in Table~\ref{tab:spectroscopic_constants}, along with data from previous theoretical calculations and experiments where available. 
    The dissociation energies \(D_e\) are estimated to be the calculated MRCI+Q energies at \(R=7.9\)\,\AA.
    The error induced by long range interactions is estimated to be less than 0.0010\,eV based on the formula and quadrupole-quadrupole coefficients given by \citet{Pattillo2018}.

    \begin{deluxetable*}{ClDDDDDDD}[hbtp]
    \tablewidth{\linewidth}
    \tablecaption{Spectroscopic constants for low-lying states of CS\label{tab:spectroscopic_constants}}
    \tablehead{
        \colhead{State}                                 & \colhead{Method}                           & 
        \multicolumn2c{\(R_e\)}                    & \multicolumn2c{\(T_e\)}              & 
        \multicolumn2c{\(\omega_e\)}               & \multicolumn2c{\(\omega_e x_e\)}     & 
        \multicolumn2c{\(B_e\)}                    & \multicolumn2c{\(10^{3}\,\alpha_e\)} & 
        \multicolumn2c{\(D_e\) \tablenotemark{a}}  \\
        \colhead{}                            & \colhead{}                     & 
        \multicolumn2c{(\(\mathrm{\AA}\))}         & \multicolumn2c{(cm\(^{-1}\))}        & 
        \multicolumn2c{(cm\(^{-1}\))}              & \multicolumn2c{(cm\(^{-1}\))}        & 
        \multicolumn2c{(cm\(^{-1}\))}              & \multicolumn2c{(cm\(^{-1}\))}        & 
        \multicolumn2c{eV}
    } 
    \decimals   
    \startdata
    X\,^1\Sigma^+    & Calc \tablenotemark{b}  & 1.540 & 0       & 1284.1 & 8.2  & 0.814 & 6.0  & 7.47                     \\
                     & Expt \tablenotemark{c}  & 1.535 & 0       & 1285.1 & 6.5  & 0.820 & 5.9  & 7.43                     \\
                     & Calc \tablenotemark{d}  & 1.533 & 0       & 1286.8 & 4.9  & 0.822 & 6.0  & 7.46                     \\
    \cline{2-16}
    a\,^3\Pi         & Calc \tablenotemark{b}  & 1.577 & 27735.5 & 1129.9 & 7.8  & 0.777 & 6.9  & 4.05                     \\
                     & Expt \tablenotemark{e}  & 1.568 & 27661.0 & 1135.1 & 7.7  & 0.785 & 7.2  & 4.00                     \\
                     & Calc \tablenotemark{d}  & 1.569 & 27721.7 & 1133.6 & 7.1  & 0.786 & 7.7  & 4.05                     \\
    \cline{2-16}
    a'\,{^3}\Sigma^+ & Calc \tablenotemark{b}  & 1.737 & 31411.7 & 832.1  & 5.9  & 0.640 & 5.5  & 3.59                     \\
                     & Expt \tablenotemark{f}  & 1.725 & 31331.4 & 830.7  & 5.0  & 0.649 & 6.0  & 3.55                     \\
                     & Calc \tablenotemark{d}  & 1.720 & 31310.2 & 829.4  & 12.9 & 0.652 & 9.1  & 3.59                     \\
    \cline{2-16}
    d\,^3\Delta      & Calc \tablenotemark{b}  & 1.753 & 35585.8 & 787.0  & 3.4  & 0.629 & 5.6  & 3.07                     \\
                     & Expt \tablenotemark{f}  & 1.742 & 35675.0 & 795.6  & 4.9  & 0.637 & 6.1  & 3.01                     \\
                     & Calc \tablenotemark{d}  & 1.741 & 35863.5 & 795.9  & 5.3  & 0.635 & 7.5  & 3.04                     \\
    \cline{2-16}
    e\,^3\Sigma^-    & Calc \tablenotemark{b}  & 1.767 & 38470.4 & 749.8  & 3.5  & 0.619 & 6.2  & 2.71                     \\
                     & Expt \tablenotemark{g}  & 1.766 & 38683.  & 752    & 4.7  & 0.619 & 4.   & 2.64                     \\
                     & Calc \tablenotemark{d}  & 1.762 & 38810.6 & 751.4  & 4.5  & 0.622 & 6.6  & 2.67                     \\
    \cline{2-16}
    A\,^1\Pi         & Calc \tablenotemark{b}  & 1.575 & 38779.4 & 1052.4 & 9.2  & 0.780 & 8.9  & 2.67                     \\
                     & Expt \tablenotemark{g}  & 1.574 & 38904.4 & 1073.4 & 10.1 & 0.780 & 6.3  & 2.61                     \\
                     & Calc \tablenotemark{d}  & 1.565 & 38943.2 & 1075.0 & 9.2  & 0.784 & 7.4  & 2.65                     \\
    \cline{2-16}
    1\,^1\Sigma^-    & Calc \tablenotemark{b}  & 1.771 & 38622.9 & 744.6  & 4.8  & 0.616 & 6.5  & 2.69                     \\
                     & Calc \tablenotemark{d}  & 1.767 & 39398.3 & 746.7  & 6.1  & 0.618 & 6.4  & 2.65                     \\
    \cline{2-16}
    1\,^1\Delta      & Calc \tablenotemark{b}  & 1.778 & 39626.9 & 718.0  & 3.9  & 0.611 & 6.8  & 2.59                     \\
                     & Calc \tablenotemark{d}  & 1.777 & 40197.7 & 723.2  & 5.7  & 0.612 & 6.6  & 2.54                     \\
    \cline{2-16}
    A'\,{^1}\Sigma^+ & Calc \tablenotemark{b}  & 1.945 & 55960.2 & 464.9  & 3.2  & 0.511 & 11.0 & 0.55                     \\
                     & Expt \tablenotemark{h}  & 1.944 & 56505   & 462.4  & 7.5  & 0.511 & 10.9 & 0.43                     \\
                     & Calc \tablenotemark{d}  & 1.958 & 57115.3 & 459.4  & 1.7  & 0.496 & 2.5  & 0.41                     \\
    \hline
    X\,^2\Sigma^+    & Calc \tablenotemark{b}  & 1.500 & 0       & 1369.8 & 9.1  & 0.859 & 6.3  & 11.34 \tablenotemark{i}  \\
    (\ce{CS^+})	     & Expt \tablenotemark{j}  & 1.492 & 0       & 1376.6 & 7.8  & 0.867 & 6.5  & 11.32 \tablenotemark{k}
    \enddata
    \tablenotemark{a}{Experimental \(D_e\) of \(X\,^1\Sigma^+\) is determined by adding energy of \(v=0\) to \(D_0^0 = 7.355\)\,eV \citep{Coppens1995}. Experimental \(D_e\) of other states are estimated by \(D_e\) (\(X\,^1\Sigma^+\)) and their \(T_e\)}
    \tablenotetext{b}{this work}
    \tablenotetext{c}{\citep{Huber1979,Mockler1955,Kewley1963,Lovas1974}}
    \tablenotetext{d}{\citep{Shi2013}}
    \tablenotetext{e}{\citep{Huber1979,Tewarson1968,Cossart1976,Taylor1972}}
    \tablenotetext{f}{\citep{Huber1979,Field1971,Cossart1976,Barrow1960}}
    \tablenotetext{g}{\citep{Huber1979,Barrow1960}}
    \tablenotetext{h}{\citep{Bell1972}}
    \tablenotetext{i}{Calculated ionization energy of CS ground \(X\) state}
    \tablenotetext{j}{\citep{Gauyacq1978}}
    \tablenotetext{k}{Experimental ionization energy of CS ground \(X\) state \citep{Coppens1995}}
    \end{deluxetable*}

    Of the 10 singlet electronic states calculated with \(A_1\) symmetry, 6 correspond to \(^1\Sigma^+\) at \(R=2.1\)\,\AA: \(X\), \(A^\prime\), \(B\), \(C\), \(5\) and \(6\,^1\Sigma^+\). 
    At \(R>2.2\)\,\AA, the \(6\,^1\Sigma^+\) is no longer among the first 10 \(A_1\) states in \(C_{2v}\). 
    Since the remaining part of its potential is still helpful to construct diabatic states, we include \(6\,^1\Sigma^+\) in Figure~\ref{fig:PECs}.
    
    The \(X\) state is deeply bound with an equilibrium internuclear distance of 1.540\,\AA\ and an estimated dissociation energy 7.47\,eV.
    The calculated spectroscopic constants, especially the vibrational constant \(\omega_e\), are in excellent agreement with experimental data (Table~\ref{tab:spectroscopic_constants}).
    Next is the \(A^\prime\) state, which is weakly bound with a much longer equilibrium bond distance, reflecting the fact that its main configuration has one electron excited from a bonding valence MO to an antibonding MO. 
    The \(B\) state has two potential wells. 
    The first, at 1.53\,\AA, lies near an avoided crossing with the \(A^\prime\) state, while the second is at 2.06\,\AA.
    The \(C\) state has its potential minimum at 1.54\,\AA, which is nearly identical to that of the \(X\) state. 
    The shape of the \(C\) state about 0.35\,eV higher than its equilibrium point is affected by avoided crossings with the \(B\) and \(5\,^1\Sigma^+\) states. 
    Following a maximum at 1.95\,\AA, a second weak potential well appears.
    For both the \(B\) and \(C\) states, the potential well near 1.50\,\AA\ corresponds to a Rydberg configuration, while the second well toward larger distances corresponds to a valence-bound electron configuration.
    A more detailed discussion of the spectroscopic properties of the \(B\) and \(C\) states can be found in section~\ref{subsec:dissc_cross_sections}.
    The \(5\) and \(6\,^1\Sigma^+\) states have complicated potential curves, possibly arising from avoided crossings with still higher states. 
 
    The five \(^1\Pi\) states are shown in Figure~\ref{fig:PECs}. 
    The \(A\,^1\Pi\) state has a calculated equilibrium internuclear distance of 1.575\,\AA\ with potential minimum of 38779.4\,cm\(^{-1}\), which is only 125\,cm\(^{-1}\) lower than the experimental value. 
    The calculated harmonic vibrational constant is 21\,cm\(^{-1}\) smaller than the experimental value of 1073.4\,cm\(^{-1}\).
    No experimental data are available to compare with the higher-lying \(^1\Pi\) states.
    Our calculations indicate that the \(2\) and \(3\,^1\Pi\) states have a prominent avoided crossing around 2.1\,\AA. 
    Finally, the \(4\) and \(5\,^1\Pi\) states lie close in energy and have unusual shapes, indicating significant Rydberg-valence mixing.
    The \(E\,^1\Pi\) state identified in the spectrum of \citet{Donovan1970} may be composed of a combination of the \(2\), \(3\), \(4\), and \(5\) adiabatic states in a diabatic representation.
    However, as discussed in more detail later, construction of such a complex diabatic state is complicated and we did not pursue this further.

    The \(1\,^1\Sigma^-\) state has a potential minimum of 38622.9\,cm\(^{-1}\) at \(R=1.771\)\,\AA. while the \(1\,^1\Delta\) state has a calculated equilibrium internuclear distance of 1.778\,\AA\ with a potential minimum of 39626.9\,cm\(^{-1}\). 
    They are almost degenerate, indicating that they share similar configurations.
    Because direct excitation from the ground state is forbidden, no experimental data are available for comparison.
    Our \(T_e\) values are about 800 and 600\,cm\(^{-1}\) lower respectively than those fitted by \citet{Shi2013} for the \(^1\Sigma^-\) and \(^1\Delta\) states, and our calculated \(R_e\) values are in good agreement with theirs.
    The \(2\,^1\Sigma^-\) and \(2\), \(3\), and \(4\,^1\Delta\) states are either unbound or very weakly bound, converging to the C (\(^1D\)) + S (\(^1D\)) atomic limit.

    The remaining electronic states of CS that correlate to one of the 4 lowest-energy atomic limits are triplet and quintet states.
    Quintet states are not involved in the photodissociation of ground-state CS and will not be discussed further.
    Triplet states, however, may play an important role in the predissociation of \(^1\Sigma^+\) states via the spin-orbit interaction as suggested by \citet{Donovan1970} even though direct excitation is forbidden from the ground state.
    The main features of the triplet state PECs from our calculations are briefly summarized here.

    The \(a^\prime\,^3\Sigma^+\) state has an equilibrium distance of 1.737\,\AA\ with a potential minimum 31411.7\,cm\(^{-1}\), which is only about 80\,cm\(^{-1}\) larger than the experimental value.
    The \(2\) and \(3\,^3\Sigma^+\) states have an avoided crossing at 1.98\,\AA, and the local maximum at 1.77\,\AA\ of the \(3\,^3\Sigma^+\) state represents a Rydberg-valence mixing.

    Next, the \(e\,^3\Sigma^-\) state has a potential minimum of 38470.4\,cm\(^{-1}\) at \(R=1.767\)\,\AA, and its calculated spectroscopic constants agree well with experimental data.
    Higher-energy \(^3\Sigma^-\) states are unbound and converge to either the C (\(^3P\)) + S (\(^1D\)) or C (\(^1D\)) + S (\(^3P\)) atomic limits.

    The \(a\,^3\Pi\) state has the potential minimum of 27735.5\,cm\(^{-1}\) at \(R=1.577\)\,\AA, and its calculated vibrational constant is 1129.9\,cm\(-1\), only 5.2\,cm\(^{-1}\) smaller than experimental value.
    An avoided crossing between the \(2\) and \(3\,^3\Pi\) states spans \(R=2.25\) to 2.50\,\AA\ and another one between the \(3\) and \(4\,^3\Pi\) states lies at \(R=1.94\)\,\AA.
    The PECs of higher \(^3\Pi\) states are close to each other, with complicated potential structures.

    Finally, the \(d\,^3\Delta\) state has an equilibrium bond length of 1.753\,\AA\ with potential energy minimum of 35585.8\,cm\(^{-1}\). 
    The fitted spectroscopic constants are in good agreement with those obtained by experiments.
    Higher \(^3\Delta\) states are either unbound or have shallow potential wells.

\begin{figure}[hbtp]
        \centering 
        \includegraphics[width=\columnwidth]{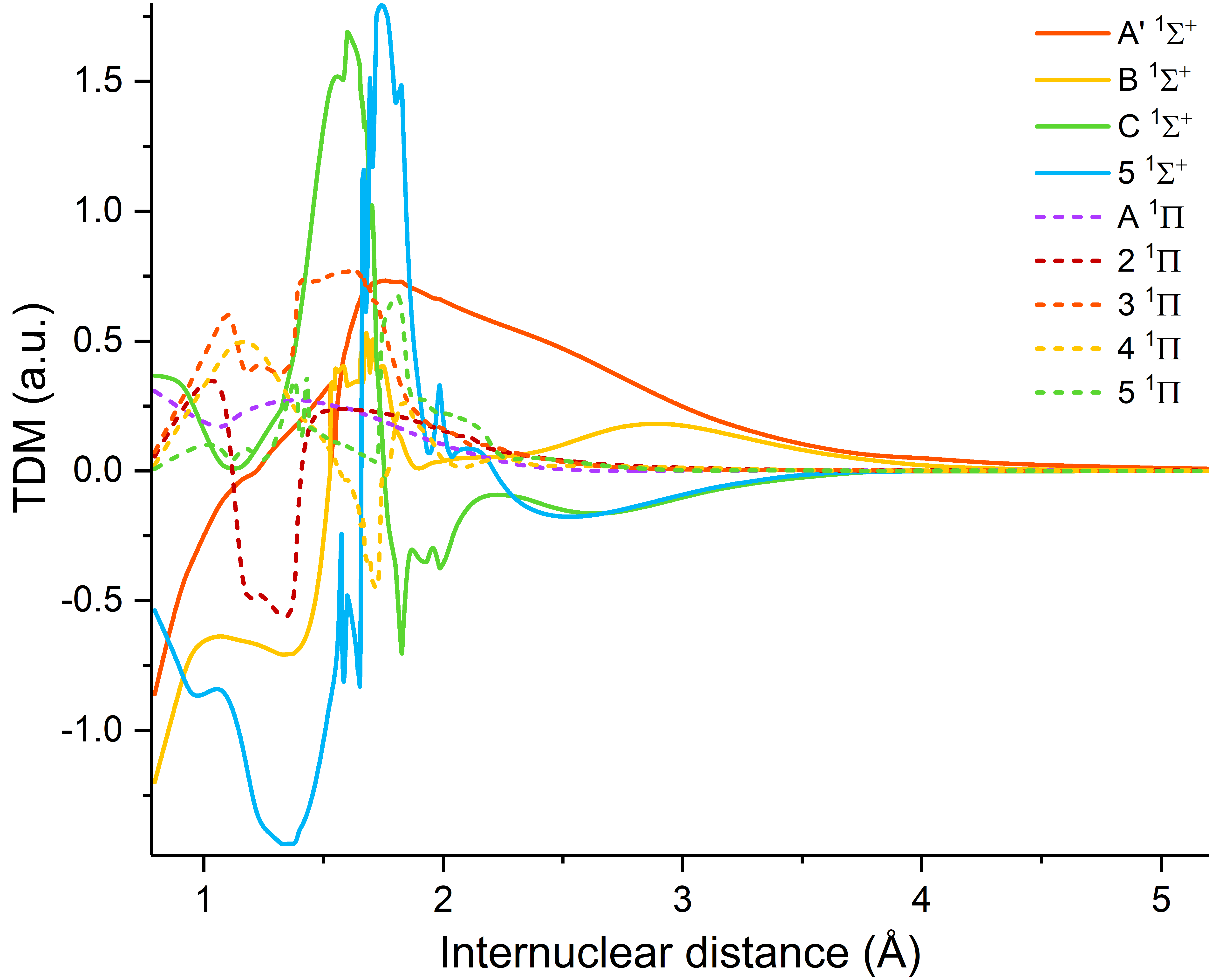}
        \caption{Transition dipole moments between the ground electronic state of CS and each excited state}\label{fig:TDMs}
    \end{figure}

    Among all these states, only those of \(^1\Sigma^+\) and \(^1\Pi\) symmetry are directly accessible from the ground \(X\) state by radiative transitions, according to the selection rules for heteronuclear diatomic molecules in Hund's case (a) and (b)
    \begin{equation}
    \Delta \Lambda = 0,\pm1; \quad \Delta S = 0; \quad + \nleftrightarrow - \label{equ:trans_selection}
    \end{equation}
    Thus, they are expected to play the most important role in the photodissociation of CS in astronomical environments.
    Transition dipole moments from the ground \(X\) state to excited \(^1\Sigma^+\) and \(^1\Pi\) states are shown in Figure~\ref{fig:TDMs}.
    We find that the \(C\) and \(5\,\Sigma^+\) states have much larger transition dipole moments compared with all other states. 
    Most importantly, the transition dipole moment of the \(C\) state is 1.5\,a.u.\ at 1.54\,\AA, which is the equilibrium internuclear distance for both the \(X\) and \(C\,^1\Sigma^+\) states.
    This indicates that the \(C-X\) \((0-0)\) transition should be extremely strong, which agrees with the experimental VUV absorption spectra of \citet{Donovan1970} and \citet{Stark1987}. 
    Among \(^1\Pi\) states, \(3\,^1\Pi\) has the largest transition dipole moment, which is about 0.7\,a.u..

    To verify our transition dipole moment calculations, we determined \(f\)-values for the \(A-X\) system for comparison with available experimental and theoretical results \citep[e.g.,][]{Carlson1979,Mahon1997,Ornellas1998,Li2013}.
    The suite of oscillator strengths is in excellent agreement.
    A selection of representative values appears in Table~\ref{tab:f_AX}. 
    
        \begin{deluxetable*}{CCCC C CCCC C CCCC}[hbtp]
            \tablewidth{\linewidth}
            \tablecaption{Calculated \(f\)-values for the \(A-X\) system\label{tab:f_AX}}
            \tablehead{
                \colhead{Band} &
                \multicolumn{3}{c}{\(f_{v^{\prime}v^{\prime\prime}}\) (\(\times 10^2\))} & &
                \colhead{Band} &
                \multicolumn{3}{c}{\(f_{v^{\prime}v^{\prime\prime}}\) (\(\times 10^2\))} & &
                \colhead{Band} &
                \multicolumn{3}{c}{\(f_{v^{\prime}v^{\prime\prime}}\) (\(\times 10^2\))} \\
                \cline{2-4}
                \cline{7-9}
                \cline{12-14}
                \colhead{(\(v^{\prime}-v^{\prime\prime}\))} &
                \colhead{This work} &
                \colhead{Expt \tablenotemark{a}} &
                \colhead{Theory \tablenotemark{b}} & &
                \colhead{(\(v^{\prime}-v^{\prime\prime}\))} &
                \colhead{This work} &
                \colhead{Expt \tablenotemark{a}} &
                \colhead{Theory \tablenotemark{b}} & &
                \colhead{(\(v^{\prime}-v^{\prime\prime}\))} &
                \colhead{This work} &
                \colhead{Expt \tablenotemark{a}} &
                \colhead{Theory \tablenotemark{b}} 
            } 
            \decimals
            \startdata
                ( 0 - 0 ) & 1.19 & 0.96    & 1.16 & & ( 2 - 0 ) & 0.02 & 0.02    & 0.03 & & ( 4 - 0 ) & 0.00 & \nodata & 0.00  \\
                ( 0 - 1 ) & 0.17 & 0.12    & 0.16 & & ( 2 - 1 ) & 0.45 & 0.35    & 0.47 & & ( 4 - 1 ) & 0.01 & \nodata & 0.01  \\
                ( 0 - 2 ) & 0.02 & 0.01    & 0.02 & & ( 2 - 2 ) & 0.40 & 0.35    & 0.34 & & ( 4 - 2 ) & 0.15 & 0.12    & 0.19  \\
                ( 0 - 3 ) & 0.00 & \nodata & 0.00 & & ( 2 - 3 ) & 0.30 & 0.24    & 0.29 & & ( 4 - 3 ) & 0.56 & 0.45    & 0.53  \\
                ( 0 - 4 ) & 0.00 & \nodata & 0.00 & & ( 2 - 4 ) & 0.08 & 0.06    & 0.08 & & ( 4 - 4 ) & 0.03 & 0.03    & 0.01  \\
                ( 0 - 5 ) & 0.00 & \nodata & 0.00 & & ( 2 - 5 ) & 0.01 & \nodata & 0.01 & & ( 4 - 5 ) & 0.23 & \nodata & 0.19  \\
                \noalign{\vskip 8pt}   
                ( 1 - 0 ) & 0.27 & 0.20    & 0.28 & & ( 3 - 0 ) & 0.00 & \nodata & 0.00 & & ( 5 - 0 ) & 0.00 & \nodata & 0.00  \\
                ( 1 - 1 ) & 0.74 & 0.63    & 0.69 & & ( 3 - 1 ) & 0.08 & 0.06    & 0.10 & & ( 5 - 1 ) & 0.00 & \nodata & 0.00  \\
                ( 1 - 2 ) & 0.27 & 0.20    & 0.26 & & ( 3 - 2 ) & 0.55 & 0.44    & 0.56 & & ( 5 - 2 ) & 0.02 & 0.02    & 0.03  \\
                ( 1 - 3 ) & 0.05 & 0.03    & 0.04 & & ( 3 - 3 ) & 0.17 & 0.15    & 0.11 & & ( 5 - 3 ) & 0.25 & 0.20    & 0.29  \\
                ( 1 - 4 ) & 0.00 & \nodata & 0.00 & & ( 3 - 4 ) & 0.29 & 0.24    & 0.26 & & ( 5 - 4 ) & 0.48 & 0.39    & 0.41  \\
                ( 1 - 5 ) & 0.00 & \nodata & 0.00 & & ( 3 - 5 ) & 0.12 & \nodata & 0.11 & & ( 5 - 5 ) & 0.00 & \nodata & 0.01
            \enddata
            \tablenotetext{a}{Derived from lifetime measurements \change{of \(R\)-branch band heads with an inherent \(\sim\)10\% uncertainty in general} \citep{Carlson1979}}
            \tablenotetext{b}{Calculated by CASSCF/MRCI \citep{Ornellas1998}}
        \end{deluxetable*}

    In their tentative detection of the CS \(C-X\) band in diffuse molecular gas, \citet{Destree2009} required estimates of the oscillator strength to derive the column density.
    To do this, they adopted an \(f\)-value of 0.14 for the \(C-X\) \((0-0)\) band based on that for the isovalent molecule CO \citep{Federman2001}.
    However, our calculations yield a significantly larger \(f\)-value of 0.45 owing to the large transition dipole moment for CS.
    With the larger CS \(f\)-value derived here, the column density inferred from the astronomical observations is substantially reduced; consequently, the expected amount of absorption of the \(A-X\) \((0-0)\) transition at 257.7\,nm (\(f = 0.096\)) is also much less, and well below the upper limit available from the astronomical measurements.

    \subsection{Coupled-channel model}\label{subsec:dissc_CSE}
    
    Several of the high-energy PECs feature avoided crossings, and in particular the \(B\) and \(C\) states are likely to share resonant levels with unbound states and therefore may decay by predissociation.
    Because the typical timescale for predissociation is much faster than spontaneous emission for small diatomic molecules, it is normally reasonable to treat their dissociation efficiency, \(\eta_d\), as unity in a collision-free environment.
    All photoabsorption is therefore expected to lead to dissociation \citep{Heays2017}.
    Experimental line broadening observed in the \(B-X\) and \(C-X\) bands supports the fast predissociation of CS \citep{Donovan1970,Stark1987}.
    In this study, we use the CSE method to investigate the predissociation of CS in detail.

    The CSE approach has been described by \citet{vanDishoeck1984} and \citet{Heays2010}.
    It has been previously used to study predissociation of other diatomic molecules, including \ce{N2} \citep{Heays2015}, \ce{O2} \citep{Gibson1996,Lewis2001}, and \ce{S2} \citep{Lewis2018}, yielding good agreement between computed and experimental cross sections. 
    In those studies, diabatic PECs are typically constructed from experimentally measured rovibrational energy levels using, for instance, the Rydberg-Klein-Rees (RKR) method.
    Then, the coupling terms and transition dipole moments are fitted iteratively by comparing the calculated cross sections, resonance positions, and widths with measured values.
    However, for many of the important predissociative states of CS, the available spectroscopic data for CS are insufficient to allow such methods. We aim to obtain the photodissociation cross sections from pure \textit{ab initio} calculations. 

    Building the coupled-channel model consists of constructing the interaction matrix \(V(R)\), whose diagonal elements are diabatic PECs and off-diagonal elements represent couplings between states.
    The wavefunctions of the coupled states are obtained from the interaction matrix and are used to derive the cross sections.
    It is impractical to include all states into the model because of the high density of states with energies above 8\,eV.
    Motivated by the strong absorption bands observed by \citet{Donovan1970} and \citet{Stark1987}, our model focuses primarily on treating predissociation originating in the \(B\) and \(C\) states. 

\begin{figure}[hbtp]
        \centering 
        \includegraphics[width=\columnwidth]{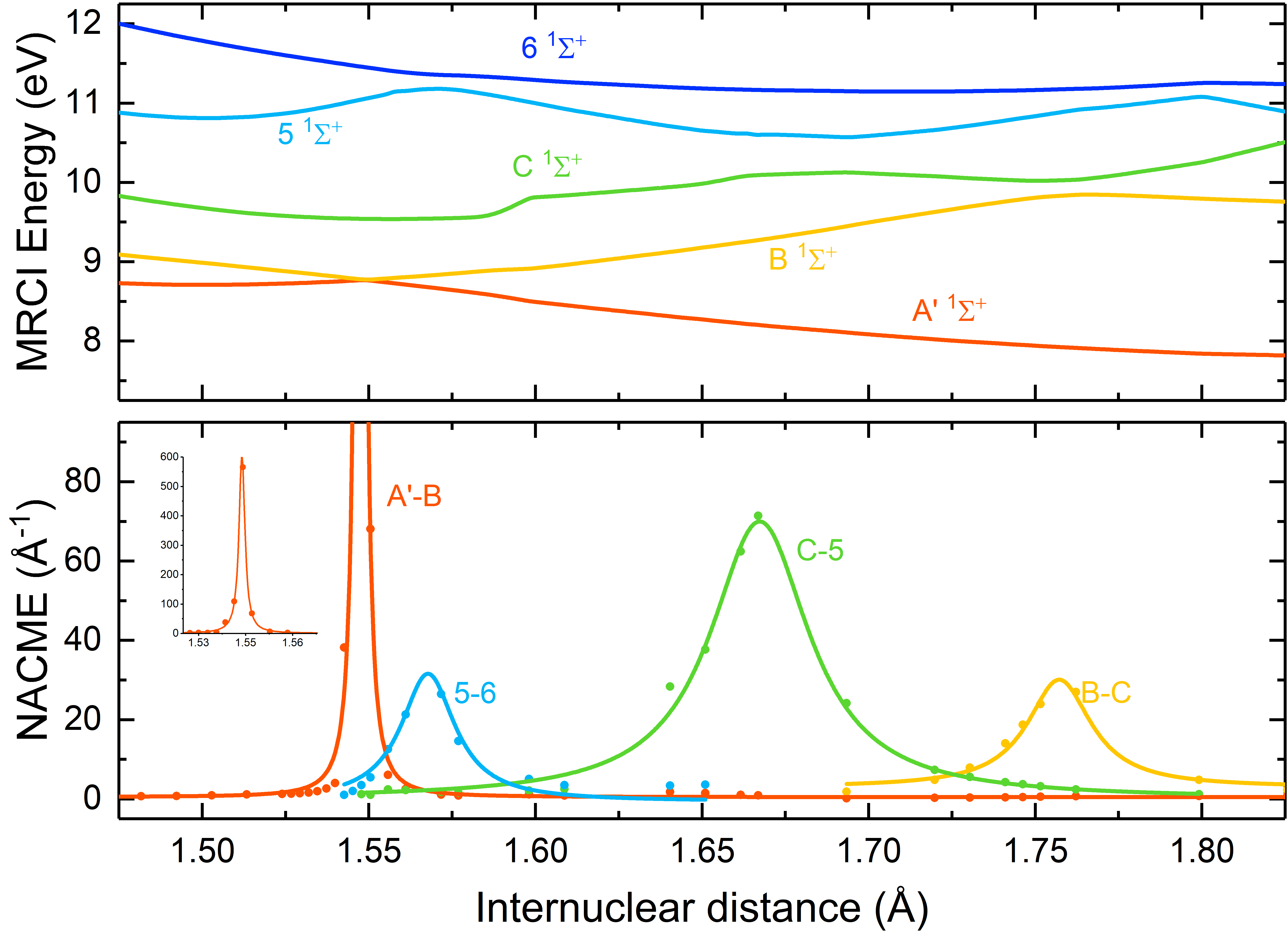}
        \caption{Adiabatic MRCI/aug-cc-5V(5+d)Z PECs for excited \(^1\Sigma^+\) states (without the Davidson correction), along with their calculated NACMEs.}\label{fig:NACME}
    \end{figure}

    We constructed diabatic PECs of the \(A'\), \(B\), \(C\), and \change{\(3^{\prime}\)}\(\,^1\Sigma^+\) states from the adiabatic ones \citep{Lefebvre-Brion2004}.
    Although it is theoretically possible to diabatize the PECs by applying the adiabatic-to-diabatic transformation matrix, which can be calculated from the non-adiabatic coupling matrix elements (NACME) \citep{Baer2006}, we did not use this method for three main reasons.
    First, calculation of the matrix at each internuclear distance requires an integral of the NACME from infinite separation.
    Small errors in each NACME may accumulate during the integration and yield artificial PECs.
    Second, solving the matrix is difficult for a system larger than two states, and the procedure is even more complicated in this case because the crossings between \(B\), \(C\), and \(5\,^1\Sigma^+\) states are close.
    Moreover, the NACMEs are calculated only at the MRCI level, corresponding to MRCI energies shown in Figure~\ref{fig:NACME}.
    The MRCI+Q energies, which include the Davidson correction, are more accurate and smoother.
    Instead, we diabatize the states by exchanging the adiabatic MRCI+Q energies on both sides of the crossing ranges and connecting the PEC segments linearly.
    For \(B\) and \(C\) states beyond our data region, we extend their PECs according to the shape of MRCI+Q PEC of \ce{CS+} calculated with the same basis set and active space.

    Initial values for the \(R\)-independent diabatic coupling matrix elements, \(H^e\), are estimated to be half the energy gaps at the crossing points shown in Figure~\ref{fig:PECs} \citep{Lefebvre-Brion2004}.
    For example, the diabatic coupling between the \(B\) and \(3^{\prime}\) states corresponds to the coupling between the \(B\) and \(C\) states in the adiabatic representation.
    The \(B\) and \(3^{\prime}\) diabatic potentials cross at 1.74\,\AA, with an MRCI+Q energy difference of 566\,cm\(^{-1}\) between \(B\) and \(C\) diabatic states, yielding an estimate for \(H^e(B,3^{\prime})\) of 283\,cm\(^{-1}\).
    
    We explored another method to estimate the diabatic coupling matrix elements without numerical integration \citep{Lefebvre-Brion2004}.
    Normally it can be assumed that the energy differences between two diabatic potentials, \(E_1^d(R)\) and \(E_2^d(R)\), vary linearly with internuclear distance \(R\) in the crossing region
    \begin{equation}\label{equ:Hel_diabatic_1}
        E_1^d(R) - E_2^d(R) = a (R-R_c)
    \end{equation}
    where \(R_c\) is the crossing point of these two potentials and \(a\) is the linear coefficient. Then the shape of the NACME forms a Lorentzian peak near \(R_c\) with a full width at half maximum (FWHM) of \(4 H^e/a\):
    \begin{equation}
        \left\langle \psi_1^{ad}\left |\frac{\partial}{\partial R} \right |\psi_2^{ad}\right\rangle_R = \frac{{H^e}/{a}}{4({H^e}/{a})^2+(R-R_c)^2}
    \end{equation}
    Thus, \(H^e\) can be estimated by:
    \begin{equation}
        H^e = \frac{a \times \text{FWHM}}{4}
    \end{equation}
    The NACMEs between adiabatic states calculated by MOLPRO do not include the Davidson correction (+Q); their values and the corresponding MRCI energies are shown in Figure~\ref{fig:NACME}.
    A higher and narrower NACME peak between adiabatic states means weaker coupling between the diabatic states.
    For the same example mentioned above, \(a\) is calculated to be 11.4\,eV/\AA \ and the FWHM is 0.036\,\AA, which results in \(H^e = 827\)\,cm\(^{-1}\).
    This number disagrees with the above value of 283\,cm\(^{-1}\).
    Instead, it matches with the half energy gap 788\,cm\(^{-1}\) at the crossing point at 1.75\,\AA\ in the MRCI PEC, shown in Figure~\ref{fig:NACME}.
    This latter approach fails because the Davidson correction contributes significantly to the energies of the excited states, especially near the avoided crossings in the MRCI calculation.
    
\begin{figure}[hbtp]
        \centering 
        \includegraphics[width=\columnwidth]{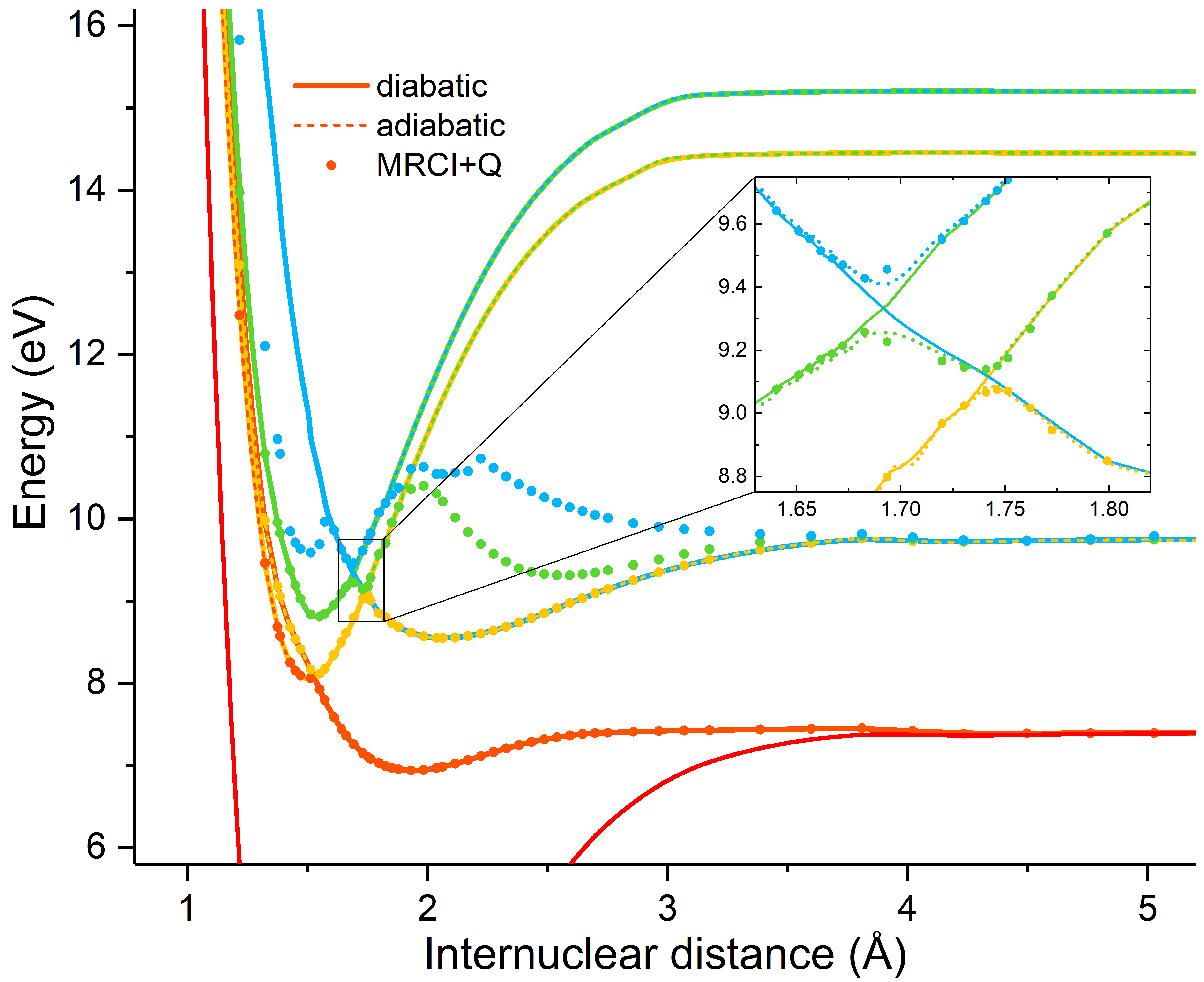}
        \caption{\(^1\Sigma^+\) states of CS: MRCI+Q \textit{ab initio} energies, constructed diabatic PECs, and their corresponding adiabatic PECs obtained by diagonalizing the diabatic interaction matrix \(V(R)\).}\label{fig:comp_cse}
    \end{figure}

    Diabatic PECs can be readily transformed back to adiabatic PECs by diagonalizing the diabatic interaction matrix \(V(R)\).
    Adiabatic PECs derived from our diabatic model in this manner should therefore agree with the calculated MRCI+Q energies.
    To improve our estimates of the diabatic state couplings, we manually refine their values to minimize the differences between the adiabatic energies derived from diagonalizing \(V(R)\) and the \textit{ab initio} energies.
    The result is shown in Figure~\ref{fig:comp_cse}.
    The perfect overlap between MRCI+Q data and adiabatic PECs validates our diabatization process.
    
    \begin{figure}[hbtp]
        \centering 
        \includegraphics[width=\columnwidth]{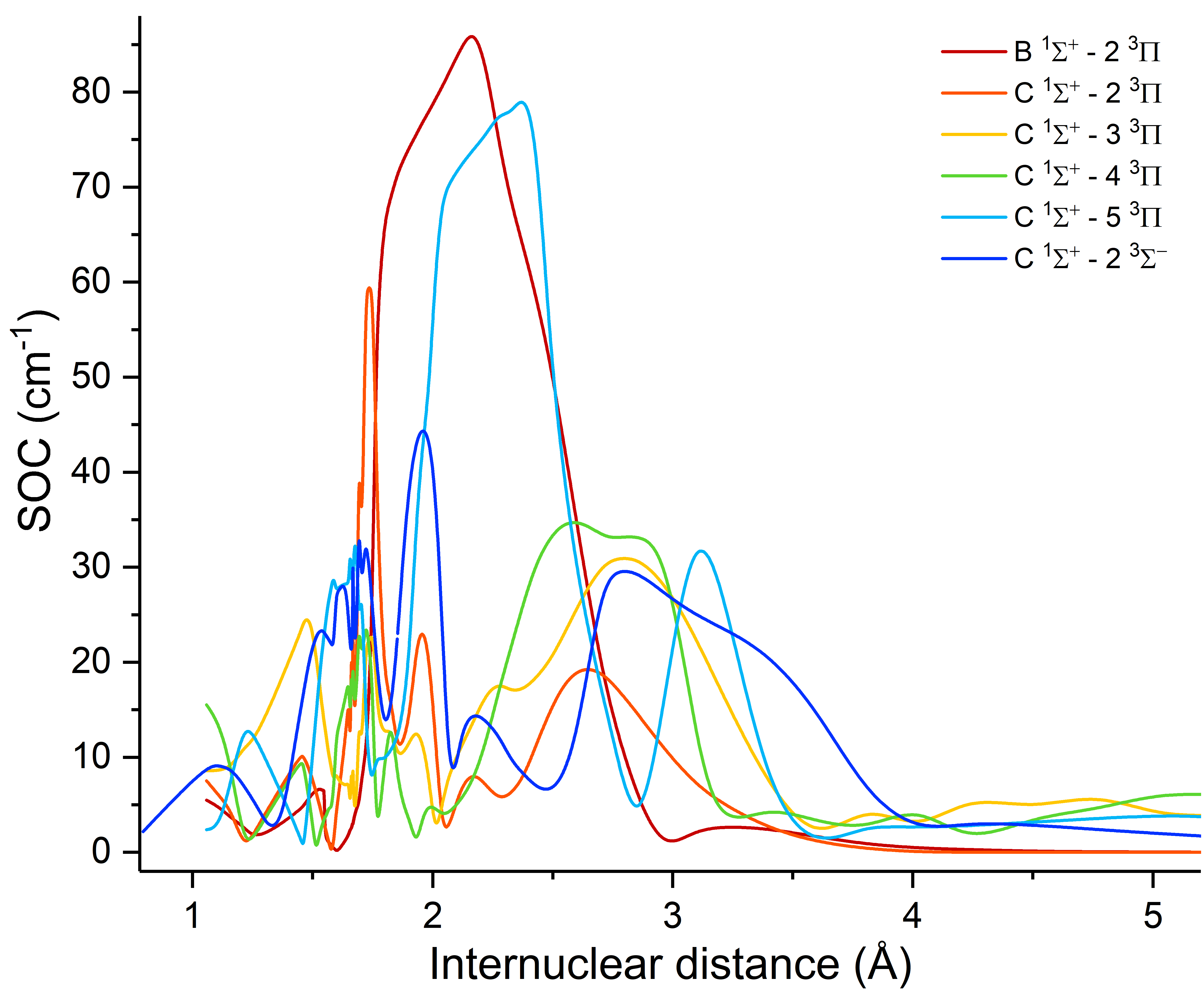}
        \caption{Several important spin-orbit couplings between \(^1\Sigma^+\) and \(^3\Pi\) states}\label{fig:SOC}
    \end{figure} 

    We also include states of \(^3\Sigma^-\) and \(^3\Pi\) symmetry in the coupled-channel model because they have non-vanishing spin-orbit coupling (SOC) terms with \(^1\Sigma^+\) states according to the selection rules:
    \begin{subequations}
        \begin{gather}
        \Delta J = \Delta \Omega = 0;\quad \Delta S = 0,\,\pm 1;\quad \Sigma^+ \leftrightarrow \Sigma^- \\
        \Delta \Lambda = \Delta \Sigma =0 \text{ or } \Delta \Lambda = -\Delta \Sigma = \pm 1 
        \end{gather}
    \end{subequations}
    In addition, obvious avoided crossings exist between the adiabatic \(2\), \(3\), and \(4\,^3\Pi\) states, as shown in Figure~\ref{fig:PECs}, and these must be included in the model.
    Diabatic PECs of \(2\), \(3\), and \(4\,^3\Pi\) states are constructed using the same method described above. 
    PECs of the higher-energy \(5\) and \(6\,^3\Pi\) states are not smooth around 1.6\,\AA, indicating strong coupling between them that is difficult to incorporate into the diabatic model.
    However, because both states are dissociative, interactions between them do not affect the overall photodissociation cross section (though they may have a small effect on the atomic product fractions), so we ignore them in our model.
    No clear crossings occur between the \(2\), \(3\) and \(4\,^3\Sigma^-\) states, so they are also treated as independent diabatic states in our coupled-channel model.

    Finally, we must include the SOCs among these diabatic states.
    Several important SOCs between the adiabatic states are shown in Figure~\ref{fig:SOC}, while all others are given in the Appendix.
    Because of the complicated adiabatic-to-diabatic transformation of \(^1\Sigma^+\) states, it is almost impossible to build \(R\)-dependent spin-orbit coupling curves for diabatic states. 
    \change{For this reason we assume that the spin-orbit interaction can be treated as \(R\)-independent, and use the values of the SOC matrix elements at the curve crossing points in the CSE model.
    This is generally a reasonable approximation when two states interact via a curve crossing.
    For instance, \citet{Lewis2018} used this approach to fit an \(R\)-independent value for the \(\left\langle 1\,^5\Pi_{u0} \right| \mathbf{H}^{\text{SO}} \left| B\,^3\Sigma_{u0}^- \right\rangle \) matrix element for \ce{S2} to experimental data, obtaining a value within 10\% of the \textit{ab initio} value calculated at the crossing point.}
    Since the crossings between \(^3\Pi\) and \(^3\Sigma^-\) are at longer internuclear distances and also above the dissociation limits, the couplings between them are unlikely to change the predissociation behavior of \(^1\Sigma^+\) states.
    We therefore do not consider the spin-orbit couplings between \(^3\Pi\) and \(^3\Sigma^-\) states in our model.
    The SOCs are calculated in MOLPRO at the MRCI level, as described in Section~\ref{subsec:theory_abinitio}.
    To convert \(\langle^1A_1|H_{\text{SO}}|^3B_1\rangle\) in the C\(_{2v}\) representation to \(\langle^1\Sigma^+|H_{\text{SO}}|^3\Pi\rangle\) in the C\(_{\infty v}\) representation, a factor of \(\sqrt{2}\) is applied.
    
    \begin{figure*}[hbtp]
        \centering 
        \includegraphics[width=\textwidth]{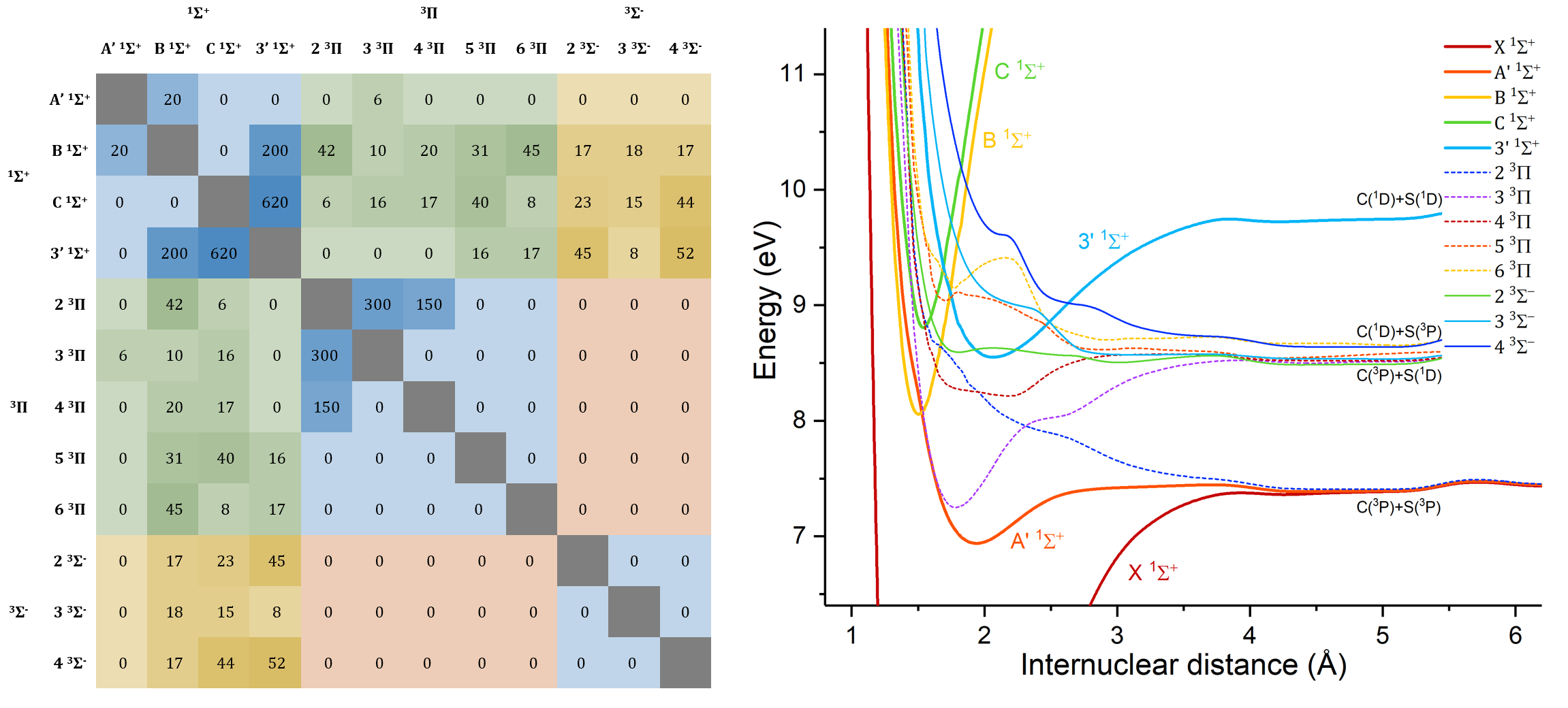}
        \caption{Coupled-channel model built for \(^1\Sigma\) diabatic states. Left: interaction matrix \(V(R)\) off-diagonal elements. Right: PECs of included diabatic states.}\label{fig:cse_model}
    \end{figure*} 

    The final potentials and coupling matrix for predissociation of the \(^1\Sigma^+\) states are shown in  Figure~\ref{fig:cse_model}.
    The diagonal elements are \(R\)-dependent potentials, while the non-diagonal elements represent the \(R\)-independent coupling terms.
    
    In addition, when calculating photodissociation cross sections for \(^1\Pi\) states, we followed a similar procedure to construct a coupled channel model for the \(2\) and \(3\,^1\Pi\) states because they have an obvious crossing at 2.15\,\AA.
    Only direct photodissociation is calculated for the \(A\), \(4\), and \(5\,^1\Pi\) states.
    
    \subsection{Photodissociation cross sections and rates}\label{subsec:dissc_cross_sections}
    
\begin{figure*}[hbtp]
    \includegraphics[width=\textwidth]{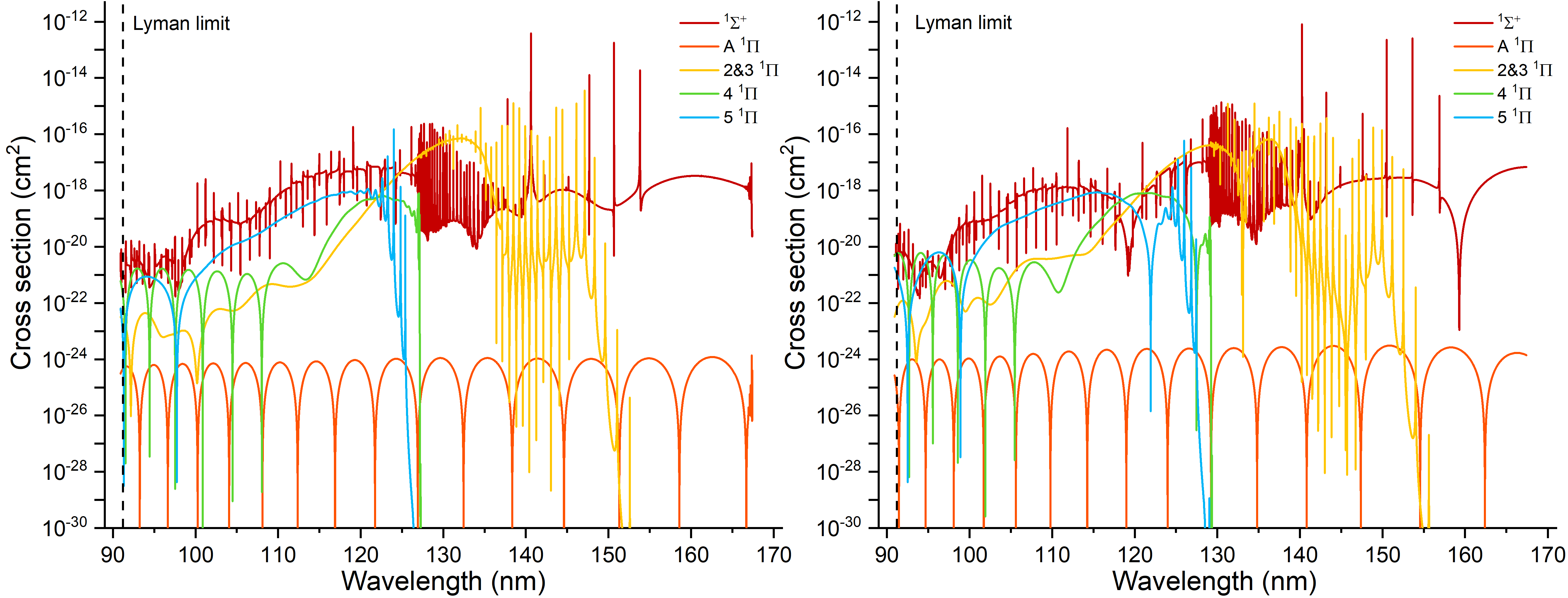}
    \caption{Rotationless photodissociation cross sections of \(^1\Sigma^+\) and \(^1\Pi\) states, \(v^{\prime\prime}=0\) (left) and \(v^{\prime\prime}=1\) (right).}\label{fig:cross_section}
\end{figure*}

    From the coupled-channel model of the \(^1\Sigma^+\) states, we calculated the rotationless photodissociation cross sections from the \(v^{\prime\prime}=0\) and \(v^{\prime\prime}=1\) vibrational levels of the ground \(X\) electronic state, without taking H\"{o}nl-London factor into consideration.
    They are shown in Figure~\ref{fig:cross_section}.
    The spectroscopic line assignments are listed in Table~\ref{tab:lines}.
    Predissociation lifetimes \(\tau_{pd}\) are calculated from the width \(\gamma\) of the peaks \citep{Kirby1989}
    \begin{equation}
    \tau_{pd} = \frac{\hbar}{\gamma} = \frac{5.3\times 10^{-12}}{\gamma}
    \end{equation}
    if \(\tau_{pd}\) is in s and \(\gamma\) is in cm\(^{-1}\).
    Spontaneous emission lifetimes \(\tau_{se}\) are derived from the inverse of the Einstein \(A\) coefficients calculated based on the integrated cross sections (\(\sigma_0\)) of those peaks.
    Those values are also included in Table~\ref{tab:lines} for comparison.
    From the calculation, we were also able to determine the dominant predissociation pathways for each transition, which are listed in Table~\ref{tab:path}.

    A comparison between the rotationless vibronic transition frequencies derived from our coupled channel model and the experimental frequencies of \citet{Donovan1970} and \citet{Stark1987} provides strong support for the accuracy of our approach.
    Our calculated \(B-X\) transition frequencies are all slightly greater than the corresponding experimental values by \(\sim150\)\,cm\(^{-1}\), while those of the \(C-X\) band are \(\sim200\)\,cm\(^{-1}\) smaller than experimental values.
    Considering the complexity of this calculation, and the fact that it is purely \textit{ab initio} with no empirical refinement, the agreement is quite satisfactory.

    Linewidths of the calculated \(B-X\) and \(C-X\) cross sections confirm their predissociative nature, especially for the origin band transitions.
    All the transitions listed in Table~\ref{tab:lines} can be treated as pure photodissociation lines in the low density conditions of the ISM because predissociation lifetimes \(\tau_{\text{pd}}\) are much smaller than the spontaneous emission lifetimes \(\tau_{\text{se}}\).
    In the experimental VUV spectrum of the \(B-X\) transition \citep{Stark1987}, only the \((1-0)\) band showed resolvable rotational structure.
    This matches our calculation, in which the \(1-0\) transition has the narrowest linewidth among the three transitions arising from \(v^{\prime\prime}=0\).
    The \(C-X\) \((0-0)\) transition has the largest cross section among all transitions from \(v^{\prime\prime}=0\) considered here by at least a factor of 30 owing to its large transition dipole moment and Franck-Condon factor.
    Its linewidth of 0.66\,cm\(^{-1}\) corresponds to \(\tau_{\text{pd}} = \) 8.0\,ps, which is over 80 times faster than \(\tau_{\text{se}}\) (0.66\,ns).

    The CSE method is also able to give the atomic product channels for each transition.
    For all \(B-X\) transitions, the dominant decay pathway is nonadiabatic coupling to the \(A'\,^1\Sigma^+\) state, leading to the ground-state C\((^3P)\) + S\((^3P)\) atomic products. 
    A small percentage (\(\sim15\%\)) couples to the \(2\,^3\Pi\) state via the spin-orbit interaction, but this also leads to the same atomic limit.
    The ground vibrational level of the \(C\) state is calculated to primarily predissociate via the \(2\), \(3\), \(4\) and \(5\,^3\Pi\) states by spin-orbit couplings. 
    Among these, the \(2\,^3\Pi\) is a minor channel corresponding to the C (\(^3P\)) + S (\(^3P\)) atomic limit, while all others (representing 89\% of the total coupling) give rise to C (\(^3P\)) + S (\(^1D\)) products.

    The photodissociation cross sections from \(^1\Pi -X\) transitions are shown in Figure~\ref{fig:cross_section}.  

\begin{deluxetable*}{CCCCCCCCCC}[hbtp]
    \tablewidth{\linewidth}
    \tablecaption{Properties of the \(B-X\) and \(C-X\) transitions of CS\label{tab:lines}}
    \tablehead{
        \multicolumn2c{Band} & 
        \colhead{\(v_{\text{expt}}\)  (cm\(^{-1}\))\tablenotemark{a}} & 
        \colhead{\(v_{\text{expt}}\) (cm\(^{-1}\)) \tablenotemark{b}} & 
        \colhead{\(v\) (cm\(^{-1}\))}                                 & 
        \colhead{\(\gamma\) (cm\(^{-1}\))}                            & 
        \colhead{\(\sigma_0\) (cm\(^2\)cm\(^{-1}\))} & 
        \colhead{\(\tau_{pd}\) (ns)}                                  & 
        \colhead{\(A\) (s\(^{-1}\))}                 & 
        \colhead{\(\tau_{se}\) (ns) }
    } 
    \decimals
    \startdata
    B-X & (0-0) & 64869\tablenotemark{c} & 64893\tablenotemark{c} & 65011.7 & 0.50 & 1.5\ee{-14} & 1.1\ee{-2} & 4.8\ee{7} & 2.1\ee{1}   \\
        & (1-0) & 66225\tablenotemark{c} & 66225\tablenotemark{c} & 66363.7 & 0.02 & 5.4\ee{-15} & 2.6\ee{-1} & 1.8\ee{7} & 5.6\ee{1}   \\
        & (2-0) & 67560\tablenotemark{c} & \nodata                & 67718.6 & 0.02 & 4.3\ee{-16} & 2.6\ee{-1} & 1.5\ee{6} & 6.7\ee{2}   \\
        & (1-1) & 64934\tablenotemark{d} & \nodata                & 65087.2 & 0.02 & 7.8\ee{-15} & 2.6\ee{-1} & 2.5\ee{7} & 4.0\ee{1}   \\
    C-X & (0-0) & 71388\tablenotemark{e} & 71327\tablenotemark{e} & 71117.7 & 0.66 & 4.0\ee{-13} & 8.0\ee{-3} & 1.5\ee{9} & 6.6\ee{-1}  \\
        & (1-0) & \nodata                & \nodata                & 72571.5 & 0.28 & 8.0\ee{-16} & 1.9\ee{-2} & 3.2\ee{6} & 3.2\ee{2}   \\
        & (1-1) & \nodata                & 71480\tablenotemark{e} & 71295.0 & 0.28 & 3.7\ee{-13} & 1.9\ee{-2} & 1.4\ee{9} & 7.1\ee{-1}
    \enddata
    \tablenotetext{a}{\citep{Stark1987}} 
    \tablenotetext{b}{\citep{Donovan1970}} 
    \tablenotetext{c}{Band origin} 
    \tablenotetext{d}{Band head position (band origin was not reported)} 
    \tablenotetext{e}{Center wavenumber of observed band}
\end{deluxetable*}

\begin{deluxetable}{CCCDC}[hbtp]
    \tablewidth{\linewidth}
    \tablecaption{Dominant predissociation pathways and product branching fractions for the \(B-X\) and \(C-X\) transitions of CS.\label{tab:path}}
    \tablehead{
        \colhead{Transition} & 
        \colhead{Band} & 
        \colhead{Channel} &
        \multicolumn2c{Percent} & 
        \colhead{Atomic products}  
    } 
    \decimals
    \startdata
        & (0-0) & A'\,^1\Sigma^+ & 85.5 & \text{C}(^3P)+\text{S}(^3P)  \\
        &       & 2\,^3\Pi       & 14.5 & \text{C}(^3P)+\text{S}(^3P)  \\
    B-X & (1-0) & A'\,^1\Sigma^+ & 96.0 & \text{C}(^3P)+\text{S}(^3P)  \\
        &       & 2\,^3\Pi       & 4.0  & \text{C}(^3P)+\text{S}(^3P)  \\
        & (2-0) & A'\,^1\Sigma^+ & 95.1 & \text{C}(^3P)+\text{S}(^3P)  \\
        &       & 2\,^3\Pi       & 4.9  & \text{C}(^3P)+\text{S}(^3P)  \\
    \hline
        & (0-0) & 4\,^3\Pi       & 48.1 & \text{C}(^3P)+\text{S}(^1D)  \\
        &       & 5\,^3\Pi       & 30.9 & \text{C}(^3P)+\text{S}(^1D)  \\
        &       & 2\,^3\Pi       & 11.0 & \text{C}(^3P)+\text{S}(^3P)  \\
        &       & 3\,^3\Pi       & 10.0 & \text{C}(^3P)+\text{S}(^1D)  \\
    C-X & (1-0) & 5\,^3\Pi       & 35.5 & \text{C}(^3P)+\text{S}(^1D)  \\
        &       & 4\,^3\Pi       & 30.2 & \text{C}(^3P)+\text{S}(^1D)  \\
        &       & 2\,^3\Sigma^-  & 15.1 & \text{C}(^3P)+\text{S}(^1D)  \\
        &       & 2\,^3\Pi       & 11.0 & \text{C}(^3P)+\text{S}(^3P)  \\
        &       & 3\,^3\Pi       & 7.7  & \text{C}(^3P)+\text{S}(^1D)
    \enddata
\end{deluxetable}

\begin{figure}[hbtp]
        \includegraphics[width=\columnwidth]{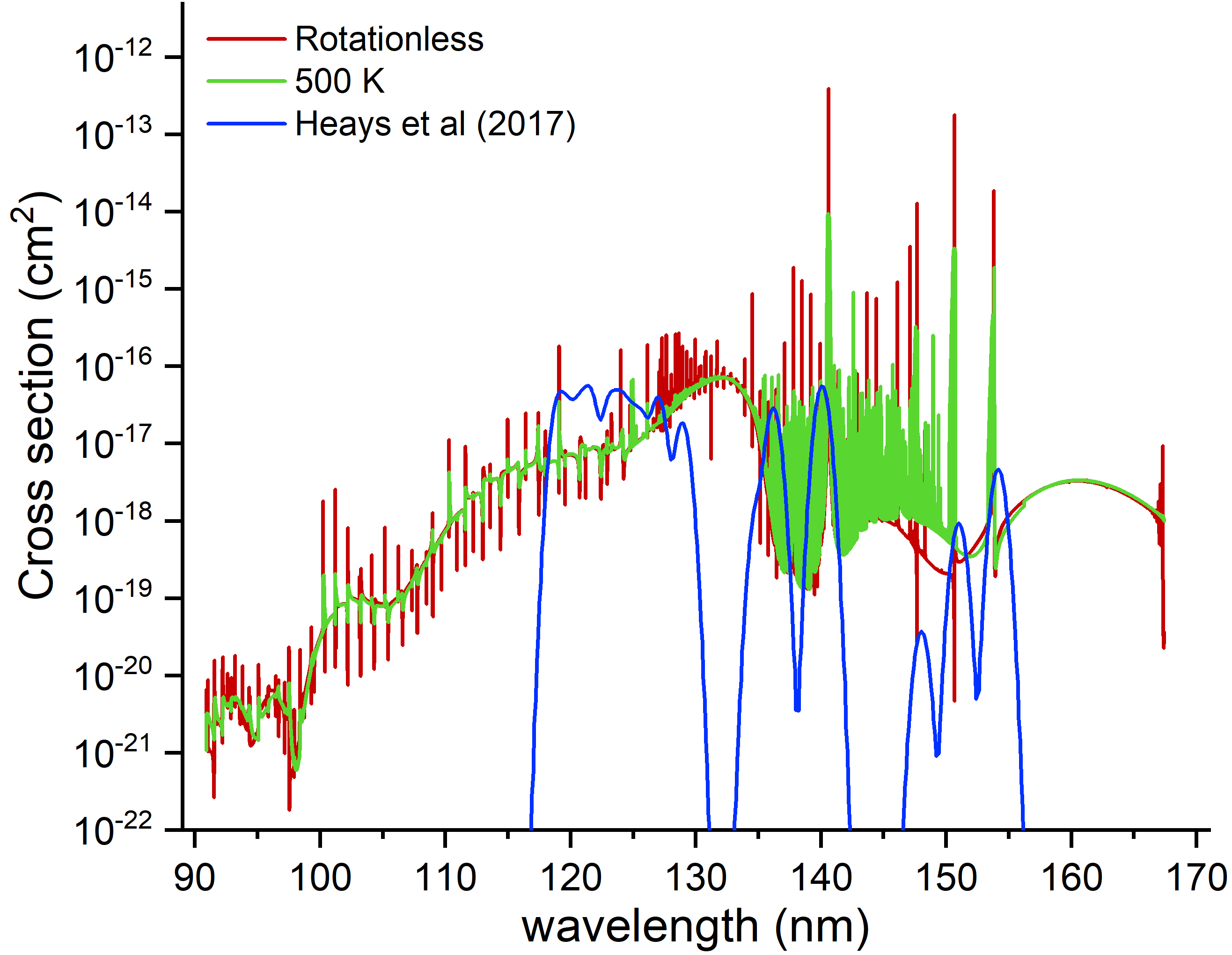}
        \caption{Photodissociation cross sections calculated at LTE temperature 500\,K, compared with data adopted in \citet{Heays2017}.}\label{fig:cross_section_lte}
    \end{figure}

    Rovibronic transitions are then calculated with applying appropriate selection rules and H\"{o}nl-London factors.
    Local thermodynamic equilibrium (LTE) cross sections are obtained using the method described in Section~\ref{subsec:theory_pd_cross_section}.
    The LTE cross sections at temperature 500\,K are shown in Figure~\ref{fig:cross_section_lte}.
    The rotational constant of the \(B(v=1)\) state is calculated to be 0.846\,cm\(^{-1}\) from the spectrum of the \(B-X\) (\(1-0\)) transition shown in the Appendix, which is almost same as 0.852\,cm\(^{-1}\) obtained from the measured spectrum \citep{Stark1987}.

    By combining the calculated cross sections and the ISRF \citep{Draine1978}, the total photodissociation rate of CS at different temperatures are shown in Table~\ref{tab:rate}, compared with the results obtained by \citet{Pattillo2018} and the standard values in the Leiden photodissociation database \citep{Heays2017}.
    The dominant transition responsible for the photodissociation of CS in space is the \(C-X\) \((0-0)\) band, comprising about 57\% of the total photodissociation in the rotationless case.
    Dissociation through \(^1\Pi\) states contribute about 32\% to the overall rate.
    Our calculated rate is a factor of 7.7 larger than that calculated by \citet{Pattillo2018} and a factor of about 3.0 larger than the value adopted by \citet{Heays2017}.

    \begin{deluxetable*}{RRCCCCCC}[hbtp]
        \tablewidth{\linewidth}
        \tablecaption{CS photodissociation rates (s\(^{-1}\)) under the standard ISRF \citep{Draine1978}.\label{tab:rate}}
        \tablehead{
            \multicolumn2c{Source} & 
            \colhead{Rotationless} &
            \colhead{20\,K} & 
            \colhead{100\,K} &
            \colhead{500\,K} &
            \colhead{\citet{Pattillo2018}} &
            \colhead{\citet{Heays2017}}
        }                   
        \decimals
        \startdata
        B-X                                  & (0-0)        & 7.03\ee{-11} & 7.11\ee{-11} & 7.10\ee{-11} & 6.82\ee{-11} & \nodata      & \nodata \\
                                             & (1-0)        & 4.50\ee{-12} & 4.41\ee{-11} & 3.43\ee{-11} & 2.61\ee{-11} & \nodata      & \nodata \\
                                             & (2-0)        & 8.71\ee{-13} & 3.95\ee{-12} & 4.15\ee{-12} & 3.01\ee{-12} & \nodata      & \nodata \\
        C-X                                  & (0-0)        & 1.64\ee{-9}  & 1.64\ee{-9}  & 1.64\ee{-9}  & 1.64\ee{-9}  & \nodata      & \nodata \\
                                             & (1-0)        & 3.24\ee{-12} & 3.24\ee{-12} & 3.29\ee{-12} & 3.58\ee{-12} & \nodata      & \nodata \\
        \multicolumn2r{\(\text{Remaining}\ ^1\Sigma^+ -X\)} & 2.33\ee{-10} & 1.95\ee{-10} & 2.18\ee{-10} & 2.39\ee{-11} & \nodata      & \nodata \\
        \multicolumn2r{All \(^1\Sigma^+ -X\)}               & 1.96\ee{-9}  & 1.96\ee{-9}  & 1.97\ee{-9}  & 1.98\ee{-9}  & 1.94\ee{-10} & \nodata \\
        \hline
        \multicolumn2r{\(A\,^1\Pi-X\)}                      & 6.57\ee{-17} & 6.57\ee{-17} & 6.61\ee{-17} & 6.88\ee{-17} & 1.50\ee{-21} & \nodata \\
        \multicolumn2r{\(2 \text{ and } 3\,^1\Pi-X\)}       & 8.84\ee{-10} & 8.84\ee{-10} & 8.85\ee{-10} & 9.07\ee{-10} & 1.35\ee{-10} & \nodata \\
        \multicolumn2r{\(4\,^1\Pi-X\)}                      & 7.14\ee{-12} & 7.14\ee{-12} & 7.11\ee{-12} & 6.96\ee{-12} & 4.05\ee{-11} & \nodata \\
        \multicolumn2r{\(5\,^1\Pi-X\)}                      & 1.54\ee{-11} & 1.55\ee{-11} & 1.58\ee{-11} & 1.62\ee{-11} & \nodata      & \nodata \\
        \multicolumn2r{All \(^1\Pi-X\)}                     & 9.07\ee{-10} & 9.06\ee{-10} & 9.08\ee{-10} & 9.30\ee{-10} & 1.76\ee{-10} & \nodata \\
        \hline
        \multicolumn2r{Total}                               & 2.86\ee{-9}  & 2.87\ee{-9}  & 2.88\ee{-9}  & 2.91\ee{-9}  & 3.70\ee{-10} & 9.49\ee{-10}
        \enddata
    \end{deluxetable*}
    
    At higher energies, \citet{Donovan1970} identified three bands at 122.93, 121.10, and 121.91\,nm, which were in turn tentatively assigned as the \(G\,^1\Pi-X\) (0-0) and (1-0) transitions as well as a forbidden transition.
    Because these transitions occur near Ly-$\alpha$, they may provide important contributions to the total CS photodissociation rate in regions where Ly-$\alpha$ is dominant.
    Our calculations do not show bands that match those reported in the 122 nm\,region.
    The calculated \(^1\Sigma^+\) states show a smooth cross section due to direct photodissociation in this region, and while the \(4\) and \(5\,^1\Pi\) states show large direct cross sections around 121.6\,nm, they are still about one order of magnitude lower than the values given in the Leiden database.
    The lack of discrete bands around 122\,nm is likely due to the limited number of states in the MRCI calculation and perhaps also due to the limited number of \(\pi\) orbitals included in the active space.
    While we were able to calculate electronic energies in this range for both \(^1\Sigma\) and \(^1\Pi\) states, the potential energy curves were not smooth and continuous.
    Therefore only direct photodissociation from lower excited states was calculated in the 122\,nm energy range. 
    Consequently, our cross sections are expected to be highly uncertain in the Ly-$\alpha$ region.
    In addition, it should be noted that the cross sections in the Leiden photodissociation database \citep{Heays2017} are also highly uncertain in this same region, so additional work is needed to address the potential importance of CS photodissociation by Ly-$\alpha$.

    \section{conclusion}\label{sec:conclu}

    Here we have presented a detailed \textit{ab initio} theoretical study of CS photodissociation from its ground electronic state using potential energy curves calculated with the MRCI+Q method with a custom basis set derived from aug-cc-pV(5+d)Z with additional diffuse functions. To improve the quality of the calculation for high-lying excited states, especially for the \(B\,^1\Sigma^+\) and \(C\,^1\Sigma^+\) states that have known strong predissociative bands from previous experiments, an expanded active space including more Rydberg MOs was used.
    Our calculation yields spectroscopic constants for the ground \(X\) and several low-lying excited electronic states in excellent agreement with experimental data.
    
    Photodissociation cross sections were calculated using coupled-channel models for excited states from the \textit{ab initio} calculation, considering both non-adiabatic and spin-orbit couplings.
    By combining these cross sections with the ISRF, CS photodissociation rates were derived at a variety of LTE temperatures along with the dominant atomic product channels.
    In space, the dominant photodissociation process for CS occurs through the \(C-X\) transition followed by spin-orbit coupling to several \(^3\Pi\) and \(^3\Sigma^-\) states, yielding C atoms in the ground \(^3P\) state and S atoms in the metastable \(^1D\) state.
    Compared with other estimates of CS photodissociation, we obtain a rate that is a factor of 2.4 larger than that adopted by the Leiden database \citep{Heays2017}.
    Our rates are about a factor of 6 greater than those estimated in the recent calculation of \citet{Pattillo2018}, arising from the fact that their choice of active space provided an inadequate treatment of Rydberg \(^1\Sigma^+\) states that have strong transitions from the ground electronic state.
    
    Finally, we would like to give an overall estimate of the accuracy of our results. 
    The foundation of our photodissociation cross sections and rates is the PECs and TDMs obtained from the \textit{ab initio} MRCI+Q calculation, which is highly reliable judged by all available spectroscopic data.
    Transition linewidths derived from the CSE calculation show that the \(B-X\) and \(C-X\) transitions can be considered completely dissociative in low-density environments where collisional relaxation is unavailable. 
    Uncertainties in the magnitudes of the couplings between states may shift the calculated transition frequencies and linewidths somewhat; however, these factors should have a minimal effect on the total calculated cross sections, which are mainly determined by the TDMs and Franck-Condon factors.
    Previous studies of diatomic molecules have shown that cross sections derived from high-level \textit{ab initio} calculations such as those employed here are generally accurate to within 20\%.
    \change{We have confidence that the cross sections calculated for the \(B-X\) and \(C-X\) bands have similar accuracy.}
    
    We found that the \(C-X\) \((0-0)\) transition is responsible for 57\% of the overall photodissociation of CS under the standard ISRF.
    This is not unexpected when compared with photodissociation of CO, as CS has a lower dissociation energy and a substantially higher density of electronic states allowing for ample opportunities for predissociation.
    While our transition frequencies differ from experimental values by about 200\,cm\(^{-1}\), as long as the radiation field is smooth in the vicinity of 140\,nm, our computed rates should be reliable.
    The atomic product branching fractions are more uncertain, as their values are sensitive to the exact methods used in the diabatization procedure.
    Future high-resolution spectroscopic measurements of the \(B-X\) and \(C-X\) bands, along with atomic branching ratios, would provide a good test for judging the ultimate accuracy of these calculations.
    They would also lead to improvements in the derived cross sections and rates because the experimentally-measured energy levels can be used to improve the diabatization and refine the \textit{ab initio} PECs. 

    Nevertheless, our calculations still have some limitations.
    First, while we proved that several low-lying vibrational states of the \(B\) and \(C\) electronic states are totally predissociative, the \change{same may not be}  true for higher vibrational states.
    Second, we are less confident in the accuracy of our electronic states at energies above the \(C\) state.
    \change{In our model, the dominant contribution to the cross section below \(\sim\)130 nm is direct photodissociation via the \(^1\,\Sigma^+\) states.
    The direct photodissocation cross section in this region for the states included in our calculation should be reliable to \(\sim\)20\%, limited primarily by the accuracy of the TDMs.
    However,} several higher \(^1\Sigma^+\) and \(^1\Pi\) states with energies below the Lyman limit exist and should also contribute somewhat to the total photodissociation, though their transition dipole moments with the ground vibronic state are likely much smaller than the \(C-X\) \((0-0)\) band.
    \change{These states will likely make a significant contribution to the total cross section via direct photodissociation, and so our calculated cross section in this region should be taken as a lower limit.
    Although any predissociation from higher-energy states would make only a small contribution to the total photodissociation rate in a smooth radiation field (much less than 10\%),} previous experiments \citep{Donovan1970} have indicated the presence of such a state near 121.6\,nm.
    Because our ab initio data and CSE calculations do not cover this energy range adequately, we do not attempt to calculate the CS photodissociation rate by Lyman-\(\alpha\) radiation, and this is an area in need of future investigation.
    
    \acknowledgments
    
    This work was supported by the NASA Astrophysics Research and Analysis program under awards 80NSSC18K0241 and 80NSSC19K0303.

    \appendix

    \section{Additional data and figures}

    \change{To verify convergence, we calculated the potential energy at several points for a series of basis sets, including aug-cc-pVQZ, aug-cc-pV5Z, aug-cc-pV6Z, aug-cc-pV(Q+d)Z with Rydberg diffuse functions, and finally aug-cc-pV(5+d)Z with Rydberg diffuse functions, which was used for the final calculations in the present study. 
    The exponents of the additional Rydberg diffuse functions are shown in Table~1.
    The potential energy curves of the \(X\), \(A'\), \(B\), \(C\), and \(A\) states are shown in Fig~10 after subtracting the energy of the \(X\) state at \(R=1.542\)\,\AA.
    Inclusion of additional Rydberg diffuse functions lowers the energy of the \(B\) state significantly.
    A maximum error of 0.08\,eV can be estimated for the \(B\) state from the difference between calculations with aug-cc-pV6Z and aug-cc-pV(5+d)Z with Rydberg diffuse functions.
    The \(A'\), \(C\), and \(A\) states are well converged.}

    \begin{figure}[hbtp]
        \includegraphics[width=\columnwidth]{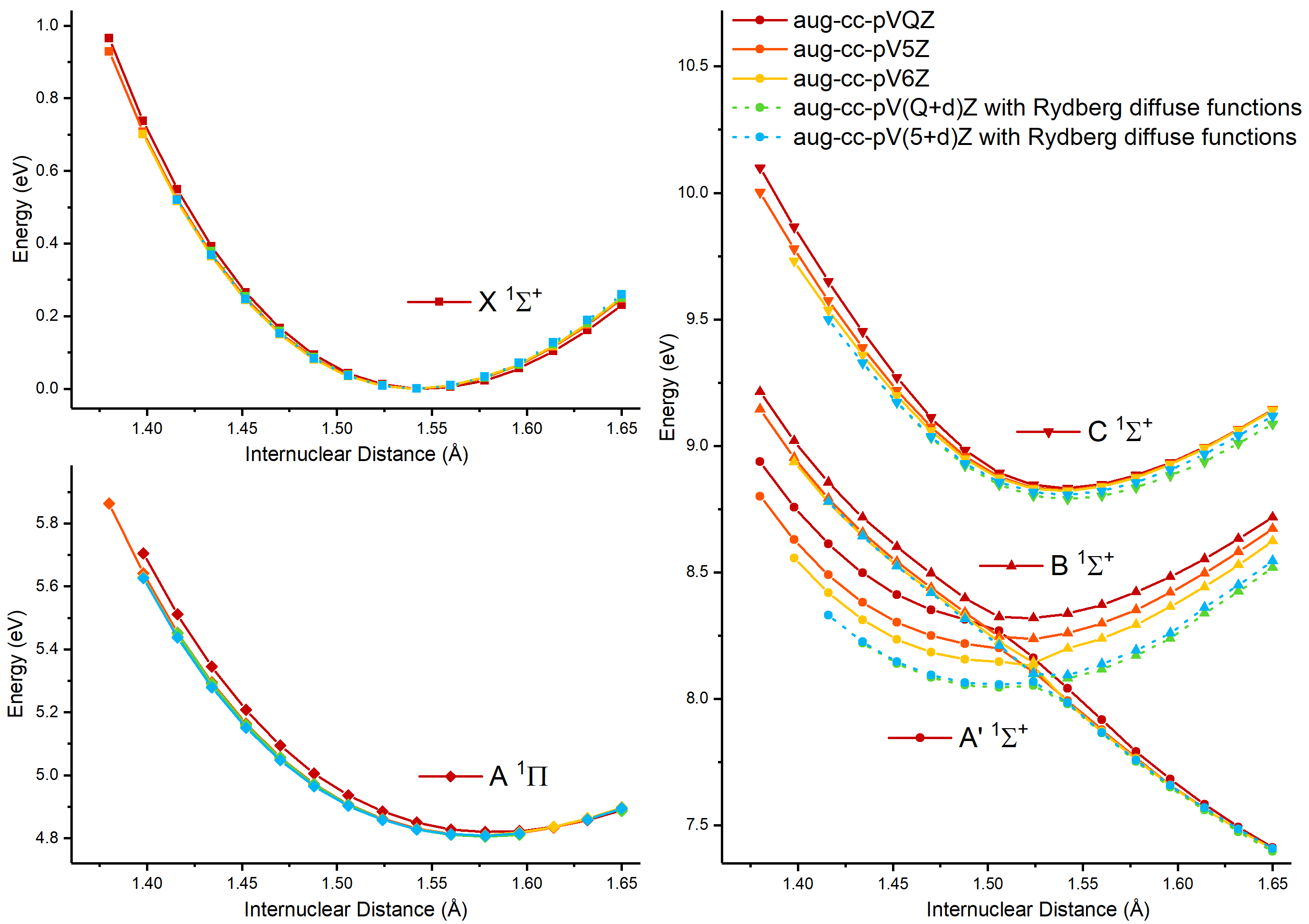}
        \caption{Potential energy curves of several states calculated with a series of basis sets}\label{fig:basis_set_convergence}
    \end{figure} 

\begin{figure*}[hbtp]
        \centering 
        \includegraphics[width=\textwidth]{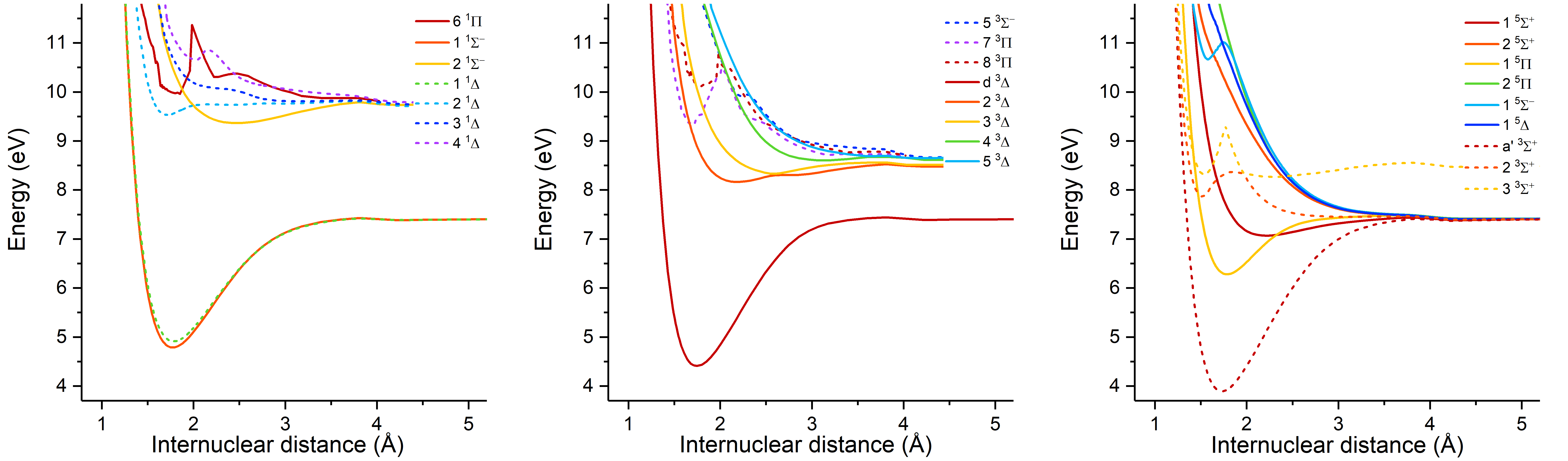}
        \caption{Potential energy curves of remaining electronic states of CS}\label{fig:pec_SI}
    \end{figure*}

    The potential energy curves of all calculated states obtained from the SA-CASSCF/MRCI+Q calculations described in Section~\ref{subsec:theory_abinitio} are given in Table~\ref{tab:app_pec}. 
    Blank entries in the table indicate that the calculation did not converge at a particular value of $R$ or that the state fell outside the range of those calculated for a particular symmetry.
    Among these 49 electronic states, the PECs of 21 states have already been shown in Figure~\ref{fig:PECs}, while the PECs of an additional 24 states are shown in Figure~\ref{fig:pec_SI}.
    The data for the remaining 4 states, \(7\,^1\Sigma^+\), \(3\,^1\Sigma^-\), \(7\,^1\Pi\), and \(4\,^3\Sigma^+\), are not shown, but the data are available in Table~\ref{tab:app_pec}.
    
    \begin{deluxetable*}{RRRRRRRRRRR}[hbtp]
        \tablewidth{\linewidth}
        \tablecaption{MRCI+Q/aug-cc-pV(5+d)Z PECs for all electronic states of CS\label{tab:app_pec}}
        \tablehead{
            \colhead{$R$ (\AA)} & 
            \colhead{$1\,^1\Sigma^+$} &
            \colhead{$2\,^1\Sigma^+$} &
            \colhead{$3\,^1\Sigma^+$} &
            \colhead{$4\,^1\Sigma^+$} &
            \colhead{$5\,^1\Sigma^+$} &
            \colhead{$6\,^1\Sigma^+$} &
            \colhead{$7\,^1\Sigma^+$} &
            \colhead{$1\,^1\Sigma^-$} &
            \colhead{\(\cdots\)} &
            \colhead{$1\,^5\Delta$}
        } 
        \decimals
        \startdata
            0.7938 & 74.87140 & 78.93278 & 79.66499 & 80.30997 & 80.72843 & 80.83307 & 86.83185 & \nodata & \cdots & \nodata   \\
            0.8996 & 43.05554 & 48.49688 & 49.24178 & 50.00352 & 50.40615 & 53.09136 & 55.71249 & \nodata & \cdots & 59.76372  \\
            0.9261 & \nodata  & \nodata  & \nodata  & \nodata  & \nodata  & \nodata  & \nodata  & \nodata & \cdots & 53.68723  \\
            0.9525 & 31.91984 & 37.79983 & 38.51802 & 39.24516 & 39.51339 & 42.98093 & 44.81884 & \nodata & \cdots & 48.29292  \\
            \multicolumn{1}{c}{\vdots} & \multicolumn{1}{c}{\vdots} & \multicolumn{1}{c}{\vdots} & \multicolumn{1}{c}{\vdots} & \multicolumn{1}{c}{\vdots} & \multicolumn{1}{c}{\vdots} & \multicolumn{1}{c}{\vdots} & \multicolumn{1}{c}{\vdots} & \multicolumn{1}{c}{\vdots} & \multicolumn{1}{c}{\vdots} & \multicolumn{1}{c}{\vdots}  \\
            7.6731 & 7.47457 & 7.48510 & 9.83764 & 9.85174 & 9.88433 & \nodata & \nodata & 7.48108 & \cdots & 7.48958  \\
            7.9377 & 7.47481 & 7.48548 & 9.83861 & 9.85228 & 9.88486 & \nodata & \nodata & 7.48135 & \cdots & 7.48967        
        \enddata
            \footnote{PECs are in eV.}
            \footnote{This table is available in its entirety in machine-readable format.}
    \end{deluxetable*}

\begin{figure*}[hbtp]
        \centering 
        \includegraphics[width=\textwidth]{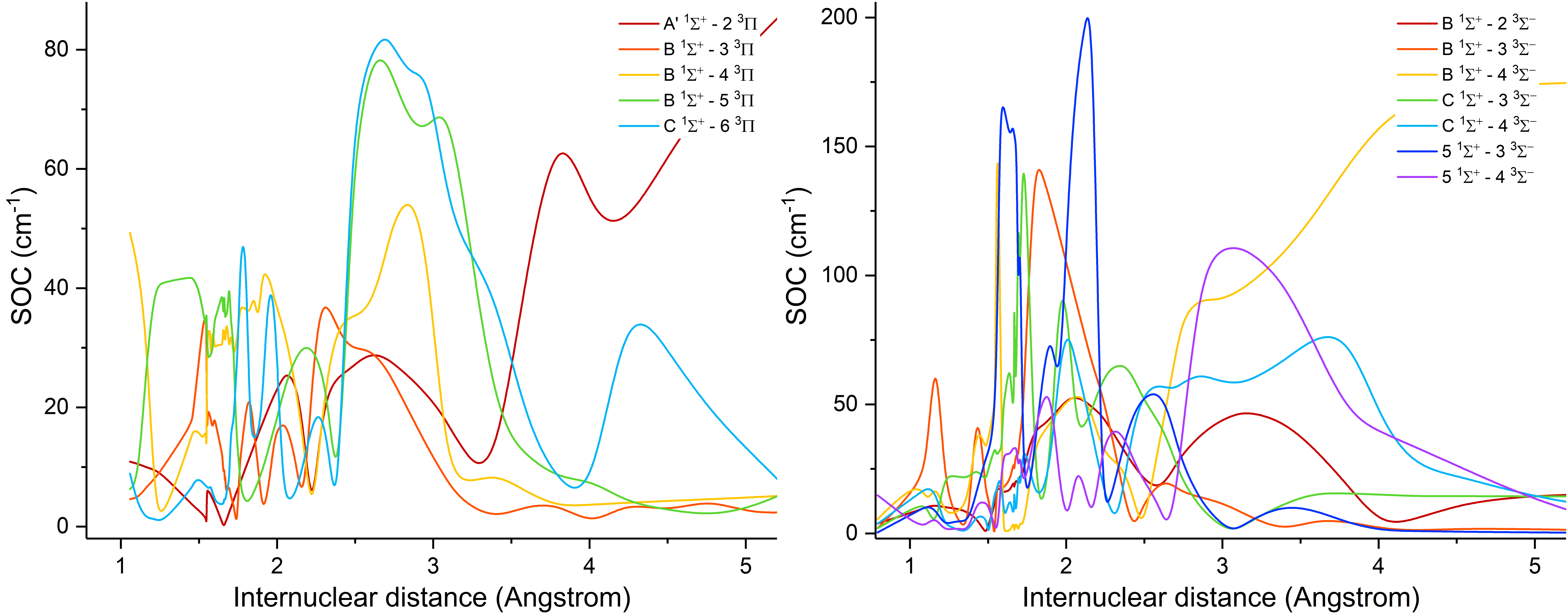}
        \caption{Several spin-orbit couplings between electronic states of CS}\label{fig:SOC_SI}
    \end{figure*}

    The absolute values of the transition dipole moments between $^1\Sigma^+$ and $^1\Pi$ states and the ground $X\,^1\Sigma^+$ state are shown in Table~\ref{tab:app_tdm}.
    Like Table~\ref{tab:app_pec}, blank entries indicate convergence failure or that the particular state was not calculated at the indicated $R$ value.
    Additionally, the spin-orbit couplings between the \(^1\Sigma^+\), \(^3\Sigma^-\) and \(^3\Pi\) states not already shown in Figure~\ref{fig:SOC} are presented in Figure~\ref{fig:SOC_SI}.
    
    \begin{deluxetable*}{RRRRRRRRRR}[hbtp]
        \tablewidth{\linewidth}
        \tablecaption{TDMs between \(^1\Sigma^+\) and \(^1\Pi\) states and the ground electronic state \(X\,^1\Sigma^+\) \label{tab:app_tdm}}
        \tablehead{
            \colhead{$R$ (\AA)} & 
            \colhead{$2\,^1\Sigma^+$} &
            \colhead{$3\,^1\Sigma^+$} &
            \colhead{$4\,^1\Sigma^+$} &
            \colhead{$5\,^1\Sigma^+$} &
            \colhead{$6\,^1\Sigma^+$} &
            \colhead{$7\,^1\Sigma^+$} & 
            \colhead{$1\,^1\Pi$} &
            \colhead{\(\cdots\)} &
            \colhead{$7\,^1\Pi$}
        } 
        \decimals
        \startdata
            0.7938 & 0.86039 & 1.19859 & 0.36731 & 0.53697 & 0.08515 & 0.19403 & 0.30781 & \cdots & 0.12908  \\
            0.8996 & 0.44859 & 0.83099 & 0.35651 & 0.75145 & 0.06795 & 0.42810 & \nodata & \cdots & \nodata  \\
            0.9525 & 0.34744 & 0.66844 & 0.26586 & 0.90374 & 0.08933 & 0.47362 & \nodata & \cdots & \nodata  \\
            1.0584 & 0.12811 & 0.63043 & 0.01976 & 0.81387 & 0.07418 & 0.52310 & 0.14462 & \cdots & 0.40322  \\
             \multicolumn{1}{c}{\vdots} & \multicolumn{1}{c}{\vdots} & \multicolumn{1}{c}{\vdots} & \multicolumn{1}{c}{\vdots} & \multicolumn{1}{c}{\vdots} & \multicolumn{1}{c}{\vdots} & \multicolumn{1}{c}{\vdots} & \multicolumn{1}{c}{\vdots} & \multicolumn{1}{c}{\vdots} & \multicolumn{1}{c}{\vdots}  \\
            7.6731 & 0.00005 & 0.00000 & 0.00000 & 0.00000 & \nodata & \nodata & 0.00039 & \cdots & \nodata  \\
            7.9377 & 0.00003 & 0.00000 & 0.00000 & 0.00000 & \nodata & \nodata & 0.00034 & \cdots & \nodata      
        \enddata	
        \footnote{TDMs are in atomic units.}
        \footnote{This table is available in its entirety in machine-readable format.}
    \end{deluxetable*}

\begin{figure*}[hbtp]
        \centering 
        \includegraphics[width=\textwidth]{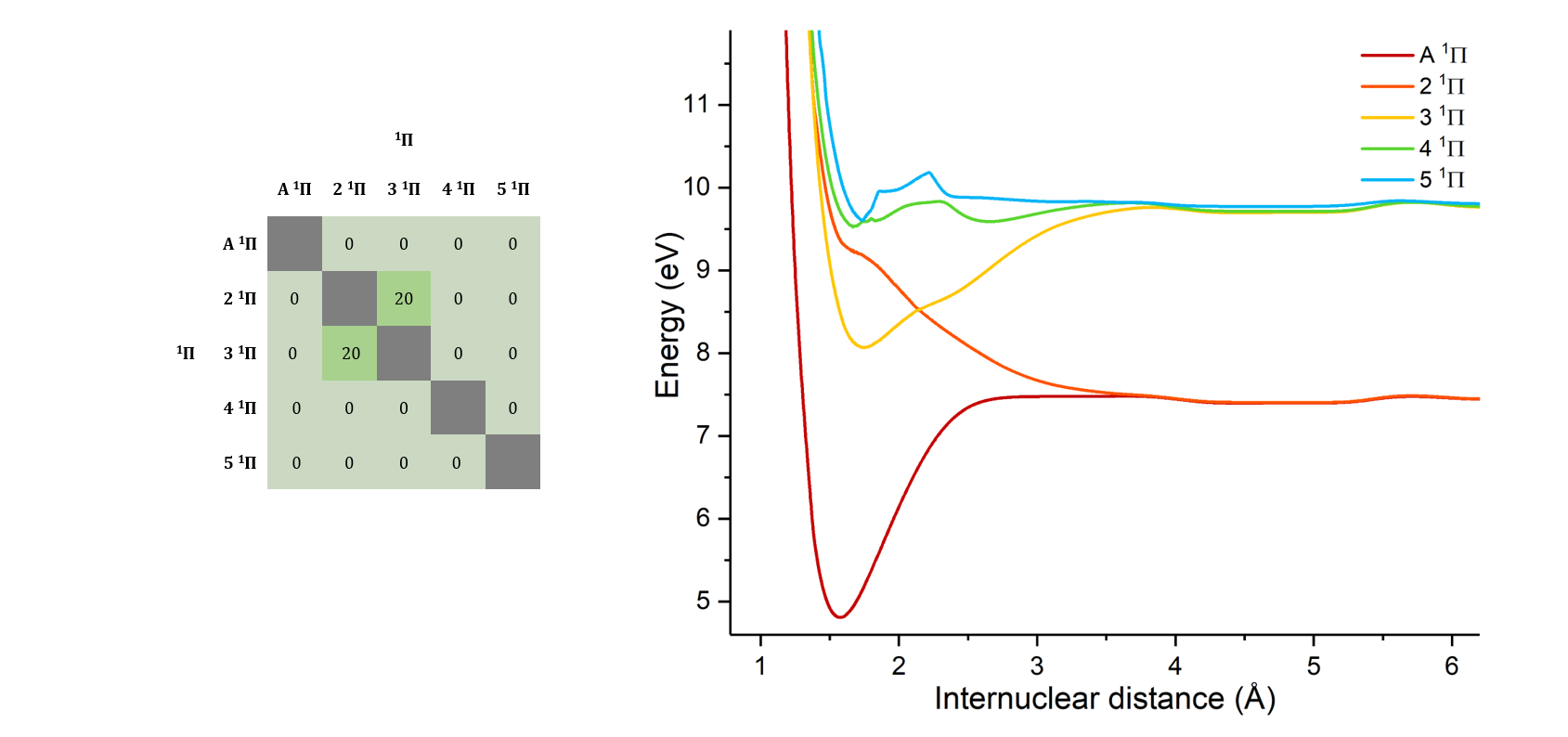}
        \caption{Coupled-channel model for the \(^1\Pi\) diabatic states. Left: interaction matrix \(V(R)\) off-diagonal elements, in cm$^{-1}$. Right: PECs of included diabatic states.}\label{fig:cse_model_pi}
    \end{figure*}

    When calculating photodissociation cross sections for \(^1\Pi\) states, we built a coupled-channel model to treat the \(2\,^1\Pi\) and \(3\,^1\Pi\) states.
    The interaction matrix is shown in Figure~\ref{fig:cse_model_pi}. 
    We ignore any couplings involving \(4\,^1\Pi\) and \(5\,^1\Pi\) states as their interactions are subtle and non-obvious.
    This may introduce errors into cross section calculations in the energy range near these states.

\begin{figure}[hbtp]
        \centering
        \includegraphics[width=\columnwidth]{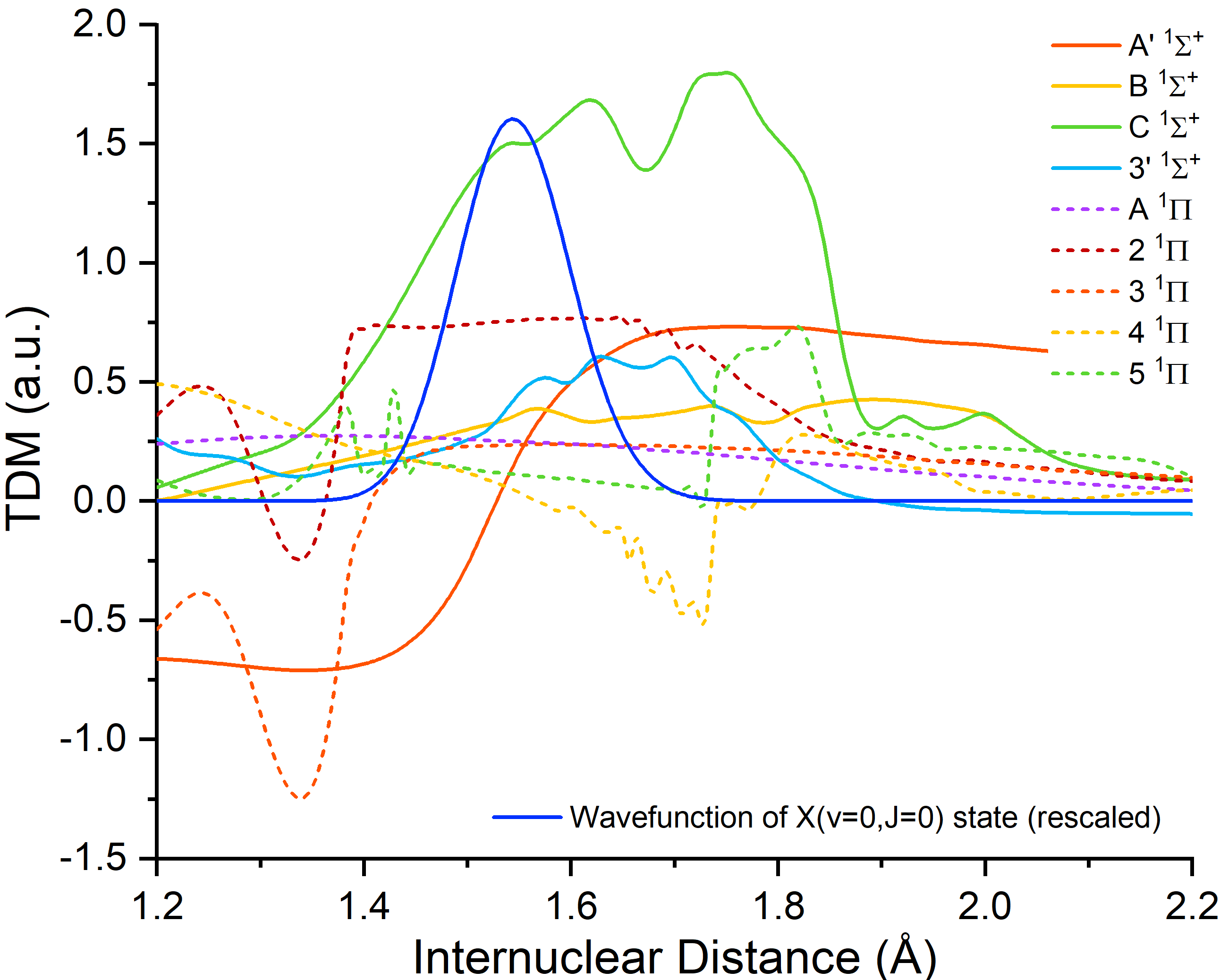}
        \caption{TDMs for diabatic states from the ground \(X\) state}\label{fig:diabatic_tdms}
    \end{figure}

    \change{The TDMs were diabatized along with the PECs by exchanging the values on both sides of the crossing points, and are shown in Figure~\ref{fig:diabatic_tdms}.
    Because diabatic PECs of the \(B\) and \(C\) states at longer internuclear distances were built by shifting PECs of the \ce{CS+} \(X\) state, we manually reduce the TDMs of these two diabatic states to 0 in this region.
    As before, we ignored any couplings involving the \(4\,^1\Pi\) and \(5\,^1\Pi\) states.
    The wavefunction of the \(X\) (\(v=0,J=0\)) state is also plotted to indicate the Franck-Condon region. 
    The diabatic PECs and corresponding TDMs used in the coupled-channel cross sections calculation are available in  Table~\ref{tab:diabatic_data}.}

    \begin{deluxetable*}{RRRRRRRRRR}[hbtp]
        \tablewidth{\linewidth}
        \tablecaption{PECs and TDMs of diabatic model built for CS \label{tab:diabatic_data}}
        \tablehead{
            \colhead{$R$ (\AA)} & 
            \colhead{pec-$X\,^1\Sigma^+$} &
            \colhead{pec-$A'\,^1\Sigma^+$} &
            \colhead{pec-$B\,^1\Sigma^+$} &
            \colhead{pec-$C\,^1\Sigma^+$} &
            \colhead{pec-$3^{\prime}\,^1\Sigma^+$} &
            \colhead{pec-$2\,^3\Pi$} & 
            \colhead{pec-$3\,^3\Pi$} &
            \colhead{\(\cdots\)} &
            \colhead{tdm-$5\,^1\Pi$}
        } 
        \decimals
        \startdata
        1.050 & 17.7367 & 24.8385 & 24.1887 & 25.8028 & 30.7668 & 30.6747 & 27.3703  & \cdots & 0.197 \\
        1.052 & 17.5116 & 24.6238 & 23.9754 & 25.5914 & 30.5403 & 30.4289 & 27.1524  & \cdots & 0.185 \\
        1.054 & 17.2886 & 24.4113 & 23.7642 & 25.3820 & 30.3157 & 30.1857 & 26.9367  & \cdots & 0.174 \\
        1.056 & 17.0677 & 24.2009 & 23.5552 & 25.1745 & 30.0930 & 29.9450 & 26.7230  & \cdots & 0.162 \\
        \multicolumn{1}{c}{\vdots} & \multicolumn{1}{c}{\vdots} & \multicolumn{1}{c}{\vdots} & \multicolumn{1}{c}{\vdots} & \multicolumn{1}{c}{\vdots} & \multicolumn{1}{c}{\vdots} & \multicolumn{1}{c}{\vdots} & \multicolumn{1}{c}{\vdots} & \multicolumn{1}{c}{\vdots} & \multicolumn{1}{c}{\vdots}  \\
        7.796 &  7.4748 &  7.4854 & 14.4318 & 15.1823 &  9.8381 &  7.4935 &  8.5675 & \cdots & 0.000 \\
        7.798 &  7.4748 &  7.4854 & 14.4318 & 15.1823 &  9.8381 &  7.4935 &  8.5675 & \cdots & 0.000
        \enddata	
        \footnote{PECs are in eV.}
        \footnote{TDMs are in atomic units.}
        \footnote{This table is available in its entirety in machine-readable format.}
    \end{deluxetable*}

    The rotationless cross sections of photodissociation from \(X\,^1\Sigma^+\) \(v^{\prime\prime}=0,1\) are given in Tables~\ref{tab:rotationless_v0} and~\ref{tab:rotationless_v1}.
    The calculation was performed with 1\,cm\(^{-1}\) resolution between 59732 and 110000\,cm\(^{-1}\), with additional points using smaller steps around several \(B-X\) and \(C-X\) transitions to better resolve the lineshapes near bound-bound transitions.
    Local thermodynamic equilibrium cross sections at various temperatures are shown in  Table~\ref{tab:lte_cs}.
    These values were calculated using Equation~\ref{equ:lte_cs} by summing over the photodissociation cross sections for the following rotational and vibrational levels the CS ground \(X\) state: \(J^{\prime\prime}=0-53\) of \(v^{\prime\prime}=0\), \(J^{\prime\prime}=0-51\) of \(v^{\prime\prime}=1\), and \(J^{\prime\prime}=0-43\) of \(v^{\prime\prime}=2\).
    These lower states were chosen so that cross sections at temperatures up to 500~K can be derived accurately.
    Owing to the increased line density at elevated temperatures, the grid size for each temperature was adapted based on the peak positions. 
    The final wavenumber grid is a collection of all the wavenumbers used.
    Cubic spline interpolation was used to generate estimated values at other small-step wavenumbers not explicitly calculated.
    
    \begin{deluxetable*}{rrrrrrr}[hbtp]
        \tablewidth{\linewidth}
        \tablecaption{Rotationless photodissociation cross sections from the \(X\,^1\Sigma^+\) \(v^{\prime\prime}=0\) level of CS \label{tab:rotationless_v0}}
        \tablehead{
            \colhead{$\tilde{\nu}$ (cm$^{-1}$)} & 
            \colhead{All $^1\Sigma^+$} &
            \colhead{$1\,^1\Pi$} & 
            \colhead{$2$ and $3\,^1\Pi$} & 
            \colhead{$4\,^1\Pi$} & 
            \colhead{$5\,^1\Pi$} & 
            \colhead{Total}
        } 
        \decimals
        \startdata
            59732.000  & 3.51e-20   & 3.10e-25   & 1.79e-96   & 0.00e+00   & 0.00e+00   & 3.51e-20    \\
            59733.000  & 3.66e-20   & 4.12e-25   & 9.93e-97   & 0.00e+00   & 0.00e+00   & 3.66e-20    \\
            59734.000  & 3.28e-20   & 7.76e-26   & 7.44e-97   & 0.00e+00   & 0.00e+00   & 3.28e-20    \\
            59735.000  & 2.96e-20   & 3.20e-26   & 6.55e-97   & 0.00e+00   & 0.00e+00   & 2.96e-20    \\
            \multicolumn{1}{c}{\vdots} & \multicolumn{1}{c}{\vdots} & \multicolumn{1}{c}{\vdots} & \multicolumn{1}{c}{\vdots} & \multicolumn{1}{c}{\vdots} & \multicolumn{1}{c}{\vdots} & \multicolumn{1}{c}{\vdots}  \\
            109998.000 & 1.75e-21   & 3.09e-25   & 3.21e-23   & 6.15e-22   & 6.38e-23   & 2.46e-21    \\
            109999.000 & 1.40e-21   & 3.08e-25   & 3.21e-23   & 6.16e-22   & 6.40e-23   & 2.11e-21
        \enddata	
        \footnote{Cross sections are in cm\(^2\).}
        \footnote{This table is available in its entirety in machine-readable format.}
    \end{deluxetable*}

    \begin{deluxetable*}{rrrrrrr}[hbtp]
        \tablewidth{\linewidth}
        \tablecaption{Rotationless photodissociation cross sections from the \(X\,^1\Sigma^+\) \(v^{\prime\prime}=1\) level of CS\label{tab:rotationless_v1}}
        \tablehead{
            \colhead{$\tilde{\nu}$ (cm$^{-1}$)} & 
            \colhead{All $^1\Sigma^+$} &
            \colhead{$1\,^1\Pi$} & 
            \colhead{$2$ and $3\,^1\Pi$} & 
            \colhead{$4\,^1\Pi$} & 
            \colhead{$5\,^1\Pi$} & 
            \colhead{Total}
        }  
        \decimals
        \startdata
            59732.000  & 6.83e-18   & 1.40e-24   & 2.19e-71   & 0.00e+00   & 0.00e+00   & 6.83e-18    \\
            59733.000  & 6.83e-18   & 1.40e-24   & 2.25e-71   & 0.00e+00   & 0.00e+00   & 6.83e-18    \\
            59734.000  & 6.83e-18   & 1.40e-24   & 2.30e-71   & 0.00e+00   & 0.00e+00   & 6.83e-18    \\
            59735.000  & 6.83e-18   & 1.40e-24   & 2.36e-71   & 0.00e+00   & 0.00e+00   & 6.83e-18    \\
            \multicolumn{1}{c}{\vdots} & \multicolumn{1}{c}{\vdots} & \multicolumn{1}{c}{\vdots} & \multicolumn{1}{c}{\vdots} & \multicolumn{1}{c}{\vdots} & \multicolumn{1}{c}{\vdots} & \multicolumn{1}{c}{\vdots}  \\
            109998.000 & 5.19e-21   & 2.72e-25   & 3.18e-23   & 4.67e-21   & 1.82e-21   & 1.17e-20    \\
            109999.000 & 5.19e-21   & 2.73e-25   & 3.18e-23   & 4.66e-21   & 1.82e-21   & 1.17e-20
        \enddata
        \footnote{Cross sections are in cm\(^2\).}
        \footnote{This table is available in its entirety in machine-readable format.}
    \end{deluxetable*}

    \begin{deluxetable*}{rrrrr}[hbtp]
        \tablewidth{\linewidth}
        \tablecaption{LTE photodissociation cross sections for ground-state CS \label{tab:lte_cs}}
        \tablehead{
            \colhead{$\tilde{\nu}$ (cm$^{-1}$)} & 
            \colhead{20K} &
            \colhead{50K} & 
            \colhead{100K} & 
            \colhead{500K} 
        } 
        \decimals
        \startdata
            59732.000  & 1.17e-19   & 4.34e-19   & 6.71e-19   & 9.99e-19    \\
            59733.000  & 1.42e-19   & 4.95e-19   & 7.24e-19   & 1.02e-18    \\
            59734.000  & 1.78e-19   & 5.88e-19   & 8.10e-19   & 1.04e-18    \\
            59735.000  & 1.91e-19   & 6.18e-19   & 8.36e-19   & 1.05e-18    \\
           \multicolumn{1}{c}{\vdots}  & \multicolumn{1}{c}{\vdots} & \multicolumn{1}{c}{\vdots} & \multicolumn{1}{c}{\vdots} & \multicolumn{1}{c}{\vdots}  \\
            109998.000 & 2.28e-21   & 1.76e-21   & 1.40e-21   & 9.78e-22    \\
            109999.000 & 1.55e-21   & 1.13e-21   & 8.74e-22   & 5.53e-22
        \enddata
        \footnote{Cross sections are in cm\(^2\).}
        \footnote{This table is available in its entirety in machine-readable format.}
    \end{deluxetable*}

\bibliographystyle{aasjournal}
\bibliography{references}

\end{document}